\documentclass[fleqn,usenatbib]{mnras}

\usepackage{newtxtext,newtxmath}

\usepackage[T1]{fontenc}

\DeclareRobustCommand{\VAN}[3]{#2}
\let\VANthebibliography\thebibliography
\def\thebibliography{\DeclareRobustCommand{\VAN}[3]{##3}\VANthebibliography}

\usepackage{graphicx}	
\usepackage{amsmath}
\usepackage{tabularx}
\usepackage{setspace}
\usepackage{natbib}

\usepackage{bm}
\usepackage{amsfonts}
\usepackage{latexsym}
\usepackage{graphicx}
\usepackage{float}
\usepackage{booktabs}
\usepackage{hyperref}
\usepackage{amsmath}
\usepackage{graphicx}
\usepackage{makecell}
\usepackage{enumitem}
\usepackage{setspace}
\usepackage{color}
\usepackage{graphicx}
\usepackage[dvipsnames]{xcolor}
\usepackage{widetext}
\usepackage[misc]{ifsym}
\hypersetup{colorlinks=true,citecolor=blue,linkcolor=red,urlcolor=magenta}
\usepackage{xcolor}
\usepackage{orcidlink}
\usepackage[caption=false]{subfig}
\title[Accelerating Universe via reconstructed H(z)]{Governing accelerating Universe via newly reconstructed Hubble parameter by employing empirical data simulations}

\author[L.Sudharani et al.]{
L. Sudharani\orcidlink{0000-0002-0860-4584},$^{1}$\thanks{E-mail: sudhaak2694@gmail.com}
Kazuharu Bamba\orcidlink{0000-0001-9720-8817},$^{2}$\thanks{E-mail: bamba@sss.fukushima-u.ac.jp}
N. S. Kavya\orcidlink{0000-0001-8561-130X}$^{1}$\thanks{E-mail: kavya.samak.10@gmail.com}
and V. Venkatesha\orcidlink{0000-0002-2799-2535}$^{1}$\thanks{E-mail: vensmath@gmail.com}\\
$^{1}$Department of P.G. Studies and Research in Mathematics, Kuvempu University, Shankaraghatta, Shivamogga 577451, Karnataka, INDIA\\
$^{2}$Faculty of Symbiotic Systems Science, Fukushima University, Fukushima 960-1296, Japan
}

\date{Accepted XXX. Received YYY; in original form ZZZ}

\pubyear{2023}

\begin{document}
\label{firstpage}
\pagerange{\pageref{firstpage}--\pageref{lastpage}}
\maketitle

\begin{abstract}
A new parametrization of the Hubble parameter is proposed to explore the issue of the cosmological landscape. The constraints on model parameters are derived through the Markov Chain Monte Carlo (MCMC) method by employing a comprehensive union of datasets such as 34 data points from cosmic chronometers (CC), 42 points from baryonic acoustic oscillations (BAO), a recently updated set of 1701 Pantheon$^+$ (P22) data points derived from Type Ia supernovae (SNeIa), and 162 data points from gamma-ray bursts (GRBs). Furthermore, the models are compared by using the Akaike Information Criterion (AIC) and Bayesian Information Criterion (BIC), so that a comparative assessment of model performance can be available. Additionally, we compare the Dainotti relation via Gaussian likelihood analysis versus new likelihoods and Calibration of the Dainotti relation through
a model-independent method. The kinematic behavior of the models is also investigated by encompassing the transition from deceleration to acceleration and the evolution of the jerk parameter. From the analysis of the parametric models, it is strongly indicated that the Universe is currently undergoing an accelerated phase with diagnostics of the model validating the quintessence phase. 
\end{abstract}




\section{Introduction}

A pivotal contemporary challenge in astronomy involves understanding the Universe's current acceleration. This cosmic speed-up is gauged primarily through the Hubble parameter value. Presently, a growing cosmic tension arises due to conflicting measurements that put forth the need for further understanding of fundamental cosmological factors. The Hubble constant $H_0$, obtained from distance ladder observations, is measured at $74.03\pm1.42~ \text{kms}^{-1}\text{Mpc}^{-1}$ \citep{Riess:2019cxk,Riess:2016jrr,Riess:2018byc}, exceeding Cosmic Microwave Background (CMB) measurements by up to 3.4$\sigma$, while from the Tip of the Red Giant (TRGB) technique (independent of Cepheid), the value of $H_0$ is $69.8\pm1.9~ \text{kms}^{-1}\text{Mpc}^{-1}$ \citep{Freedman:2019jwv} that is in agreement with Plank data and SH0ES calibrations at 1.2$\sigma$ and 1.7$\sigma$ respectively. However, using the local distance ladder with Cepheids and type Ia supernovae, the SH0ES team determines $H_0$ as $73.04 \pm1.04$ $\text{kms}^{-1}\text{Mpc}^{-1}$ \citep{Riess:2021jrx}, creating an approximately 5$\sigma$ discrepancy. Notably, a significant conflict emerges concerning the $H_0$, which gauges the Universe's expansion rate. Presently, a tension at the 5-6 $\sigma$ level exists between local measurements of $H_0$ and those from CMB models \citep{Riess:2020fzl,DiValentino:2021izs, DiValentino:2020zio}. Computed values from H0LiCOW \citep{Bonvin:2016crt,Birrer:2018vtm}, Gravitational Wave (GW) \citep{Gayathri:2020mra}, and Megamaser Cosmology Project (MCP) \citep{Pesce:2020xfe} present a higher $H_0$ range, aligned with distance ladder estimates. 

In exploring diverse astronomical observations, the $\Lambda$CDM model has demonstrated remarkable success, spanning from the early stages of the Universe, including the epoch of Big Bang nucleosynthesis. By employing mere free parameters, this predictive model unveils the fundamental dynamics of our vast cosmic scale. However, despite its success, the $\Lambda$CDM model falls short in addressing the challenge posed by conflicting measurements of primary cosmological factors \citep{Abdalla:2022yfr}, necessitating exploration beyond its confines \citep{Riess:2019cxk}.

In this context, the reconstruction methodology emerges as a promising avenue to address cosmological observations. It utilizes rigorous statistical techniques to establish a viable kinematic model. Recent attention has been directed toward this approach due to its ability to effectively align theoretical models with empirical observations, all while maintaining independence from the underlying gravity model. There are two variations of reconstruction: parametric and non-parametric. Non-parametric reconstruction involves directly deriving models from observational data using statistical procedures, while parametric reconstruction begins by defining a kinematic model with independent parameters and subsequently determining restricted redshift range through statistical analysis of observational data. This methodical process provides a compelling way to conceptually address the limitations inherent in the standard model, encompassing phenomena such as late-time acceleration, the cosmological constant problem, and the initial singularity. Through the utilization of this method, one can endeavor to enrich our understanding of the Universe, potentially untangling these core enigmas surrounding acceleration and dark energy \citep{Copeland:2006wr,Padmanabhan:2007xy,Durrer:2007re,Bamba:2012cp}.

In the study of an article \citep{Mukherjee:2016trt}, the authors explored the Universe's accelerating nature using parametric reconstruction, with a specific focus on the jerk parameter. Likewise, the trajectory of the Hubble parameter's evolution can be illuminated through the deceleration parameter (as explained by Campo et al. \citep{delCampo:2012ya}, allowing for precise predictions of thermal equilibrium. Gong et al. \citep{Gong:2006gs} further extended this parametrization technique to probe the equation of state (EoS) for dark energy. In the existing literature, numerous investigations revolve around parameterizing deceleration, jerk, and equation of state (EoS) attributes \citep{Roy:2022fif,Boughezal:2016wmq,Mukherjee:2016eqj,Pantazis:2016nky,Jaime:2018ftn,Nair:2011tg,Akarsu:2013lya,Naik:2023yhl,Naik:2023gma,Roman-Garza:2018cxf,Koussour:2023rmm,Koussour:2022jyk,Arora:2023dvs,Chaudhary:2024bop}. However, limited attention has been directed toward parameterizing the Hubble parameter \citep{Pacif:2020hai,Pacif:2016ptv}, despite its fundamental significance in elucidating the Universe's evolution. Given this, we propose a parametric formulation for the Hubble parameter, incorporating constraints on the Hubble constant as well. Our principal aim with this approach is to reconstruct a viable model that accommodates the accelerating Universe. Additionally, we examine the solutions of Einstein's field equations within the framework of an isotropic and homogeneous spacetime.

For this endeavor, we leverage statistical analysis of observational data to ascertain the constrained values governing the model's parameters. Through the utilization of the Bayesian approach in agreement with Markov Chain Monte Carlo (MCMC) analysis, we seek out the best fits for the unconstrained parameters. In our examination, we harness diverse datasets, encompassing Cosmic Chronometers (CC), Baryonic Acoustic Oscillations (BAO), Gamma-Ray Bursts (GRBs), and the recently released SH0ES (P22) dataset. This expansive compilation includes a broad spectrum of observations, spanning from Hubble Space Telescope (HST), Supernova Cosmology Project (SCP), Great Observatories Origins Deep Survey (GOODS), Sloan Digital Sky Survey (SDSS), Panoramic Survey Telescope and Rapid Response System (Pan-STRARRS), All-Sky Automated Survey for Supernovae (ASASSN), Swift Gamma-Ray Burst Mission (SWIFT), Lick Observatory Supernova Search (LOSS), Carnegie Supernova Project (CSP), Hubble Frontier Fields (HF), and more.

The layout of the article is as follows: 
Section~\ref{flrw} presents the fundamental aspects of Friedmann-Lema\^{i}tre Robertson-Walker and also proposes the novel Hubble parameterization. Based on this, Section~\ref{sec:data} provides a rigorous examination of the statistical analysis applied to the datasets. In Section~\ref{model}, we undertake a comparative evaluation of our model utilizing the Bayesian model comparison technique. Further comparative study has to be done in Section~\ref{dainotti} with the  Dainotti relation via Gaussian likelihood analysis versus new likelihoods and calibration of the Dainotti relation through a model-independent method. The subsequent section~\ref{kinematics} employs the outcomes from the statistical approach to decipher the Universe's behavior within our newly constructed model. This section also explores the physical implications of the parameterized models. Further, section~\ref{diagnize} explores the $Om$ diagnostics of the model. Lastly, our findings are summarized and concluded in Section~\ref{conclusion}.

\section{ The FLRW Spacetime: Unveiling the Generalized $\Lambda$CDM Cosmology}\label{flrw}

In the realm of cosmology, the flat Friedmann-Lema\^itre-Robertson-Walker (FLRW) metric assumes a pivotal role, depicting a cosmos that exhibits isotropy (uniformity in all directions) and homogeneity (uniformity at every locale).

Within the framework of a flat FLRW spacetime, the metric takes the following expression:

\begin{equation}
ds^2 = -dt^2 + \mathcal{A}^2(t)(dx^2 + dy^2 + dz^2).
\end{equation}

In this context, the quantity represented by $ds^2$ signifies the spacetime interval, while $dt$ corresponds to an infinitesimal increment of cosmic time. The function $\mathcal{A}(t)$ denotes the scale factor, a fundamental concept encapsulating the Universe's proportions as they evolve through time, fairly explaining the expansion of the Universe.

Central to this concept is the association of the scale factor with the Hubble parameter ($H$), which characterizes the pace at which the Universe expands. The Hubble parameter ($H$) finds its definition as the temporal derivative of the scale factor divided by the scale factor itself: $H = \frac{\dot{\mathcal{A}}}{\mathcal{A}}$, wherein the dot symbolizes differentiation with respect to cosmic time.

Upon the incorporation of the flat FLRW metric  into Einstein's field equations for a perfect fluid endowed with energy density $\rho$ and pressure $p$, it yields the Friedmann equations characterizing a spatially flat FLRW Universe

\begin{gather}
    3H^2 = \rho, \label{EQ:Friedmann01}\\
    2\dot{H} + 3H^2 = -p. \label{EQ:Friedmann02}
\end{gather}

\par In this groundwork, adhering to the convention $\frac{8\pi G}{c^2} = 1$, we encounter the equation of state (EoS) parameter ($w$), which emerges as the quotient of pressure to energy density, denoted as $w = \frac{p}{\rho}$. This parameter exerts influence over the dynamics of energy density as well as the expansive tendencies of the Universe. Notably, a Universe experiences acceleration when this parameter takes on values less than $-1/3$.

Moreover, the EoS parameter proves invaluable in categorizing distinct phases of cosmic expansion: When $w = 0$, it corresponds to the epoch characterized by matter domination. A value of $w$ less than -1 signifies the phantom phase, while $w$ values residing between $-1$ and $-1/3$ pertain to the quintessence phase.

The deceleration parameter $q$  emerges as another pivotal factor characterizing the evolutionary trajectory of the Universe. Its connection to the second time derivative of the scale factor within the FLRW metric is articulated as follows $q = -\frac{\ddot{\mathcal{A}}}{H^2\mathcal{A}}$. The deceleration parameter serves as a gauge of whether the Universe's expansion rate is either accelerating or decelerating. A positive value of $q$ signifies deceleration, while a negative value denotes acceleration.

The relation between the Hubble parameter and the deceleration parameter is encapsulated within the subsequent relationship

\begin{equation}
H(z) = H_0 \exp \left(\int_0^z [q(\kappa) + 1] d\ln (\kappa + 1)\right).
\end{equation}
Here, $H_0$ denotes the Hubble parameter's value at $z=0$, while $z$ represents the redshift, a value correlated with the scale factor `$\mathcal{A}$' through $z = -1 + \frac{1}{\mathcal{A}}$. This association furnishes a means to ascertain the Hubble parameter as a function of redshift, predicated upon the deceleration parameter.

The Hubble parameter (HP) serves as a pivotal factor in unraveling the evolution and characteristics of the Universe. Diverse physical theories and cosmological models have been proposed to elucidate its behavior. There are various physical arguments and cosmological models that explain the behavior of the Hubble parameter. However, a model-independent methodology has been proposed by \citep{Shafieloo:2012ms}. This methodology employs a cosmological parametrization and tackles the field equations with three unknowns: density, pressure, and the Hubble parameter. Cosmologists have proposed different parametrizations of cosmological parameters like  HP, $q$, and EoS. These endeavors aim to enhance our comprehension of observed phenomena in the Universe,  such as the transition from deceleration to acceleration. These model parameters can be constrained through observational data, making parametrization a useful tool for scrutinizing the Universe. In this work, we utilize parametric reconstruction of the model to investigate the accelerating expansion of the Universe. Specifically, we construct a new parametric form of the HP which is defined as 
 
\begin{equation}
    H(z)=H_0\,\left[1-a(1+z)^2+(z+a)(1+\sqrt{\Omega_M}z)^2\right]^{1/2},\label{umodel}
\end{equation}

where $a$, and $\sqrt{\Omega_M}$  are parameters to be constrained by observational data, and the value of $H_0$ corresponds to the Hubble constant. Over considering the standard $\Lambda$CDM model
\begin{equation}
    h(z)=\left[\Omega_M\,(1+z)^3+\Omega_{\Lambda}\right]^{1/2},\label{standard}
\end{equation}
 
where $h(z)=\frac{H(z)}{H_0(z)}$, $\Omega_M$ and $\Omega_{\Lambda}$ are the matter density and cosmological constant density parameters respectively. Notably, the provided Hubble parameter introduces additional parameters $a$ and $\Omega_M$ that extend the standard model. By introducing these parameters, the equation implies a more complex expansion behavior that could potentially provide a better fit to observed data. Hence our model performs well across diverse observational data, which strengthens its validity. 

The term $1-a(1+z)^2$ becomes less sensitive to the magnitude of the model parameter $a$. Meanwhile, the term $(z+a)(1+\sqrt{\Omega_M}z)^2$ shows a high sensitivity to the variations of both parameters, thereby displaying a more pronounced dependence on model parameters, which introduces a more complex dependence on redshift. The combination of the linear and quadratic terms suggests that this part of the expression might have a significant impact on expansion behavior.

By comparing the parametrized $H(z)$ given by \eqref{umodel} and with the standard expression \eqref{standard}, and solving for the model parameters $a$ and $\Omega_M$, we deduce the $\Lambda$CDM equivalent of the parametrized H(z). For this specific case, the values $a=\frac{3\Omega_M-2\sqrt{\Omega_M}}{\Omega_M-1}$ yield the $\Lambda$CDM equivalent of the parametrized H(z). In particular, parameter $\sqrt{\Omega_M}$ in the parametrized model is explicitly related to the matter density $\Omega_M$ in the $\Lambda$CDM that causes an accelerated cosmic expansion and parameter $a$ indirectly impacts the matter density by influencing the overall expansion behavior provides a different mechanism for cosmic acceleration, it could present an interesting alternative to the cosmological constant in $\Lambda$CDM. By analyzing how the parametrized model's behavior changes with variations in $\Omega_M$, we can gain a deeper understanding of how matter density affects the expansion rate and growth of the Universe according to this modified model. This exploration could lead to insights into phenomena such as the transition from deceleration to acceleration, the behavior of large-scale structure formation, and the overall dynamics of spacetime.

\par The deceleration parameter, $q$ is a vital cosmological parameter that characterizes the Universe's evolution from deceleration to acceleration. This transition is marked by positive and negative values of $q$ corresponding to early and late-time acceleration, respectively. This article employs observational data to predict the Universe's current state. The relationship between $q$ and the Hubble parameter $H$ is given by
\begin{equation}
q(z) = -\frac{\dot{H}}{H^2} - 1. \label{eq:7}
\end{equation}
A connection between cosmic time $t$ and redshift $z$ is established
\begin{equation}
\frac{d}{dt} = -(z+1)H(z) \frac{d}{dz}. \label{eq:8}
\end{equation}
By utilizing $\dot H=-(z+1)H(z)\frac{dH}{dz}$ along with equations \eqref{model}, \eqref{eq:7}, and \eqref{eq:8}  the $q$ for the model can be computed as follows

\begin{equation}
q(z)=-\frac{(-z-1) \left(2 \sqrt{\Omega_M} (a+z) (\sqrt{\Omega_M} z+1)-2 a (z+1)+(\sqrt{\Omega_M} z+1)^2\right)}{2 \left((a+z) (\sqrt{\Omega_M} z+1)^2-a (z+1)^2+1\right)}-1. \label{eq:9} 
\end{equation}
This equation provides an expression for $q$ in terms of the model's parameters $a$ and $\Omega_M$.
 
\section{Observational data set's Analyses and Results}\label{sec:data}
 In this section, we emphasize the importance of observational cosmology in developing precise cosmological frameworks. To attain this goal, it becomes of utmost importance to constrain the model parameters, specifically $a$, $\sqrt{\Omega_M}$, and $H_0$, through absolute scrutiny of observed information. In this investigation, we make use of multiple sets of observational data, encompassing Cosmic Chronometers (CC), Baryonic Acoustic Oscillations (BAO), and the most recent Pantheon$^+$ compilation (P22), derived from studying Type Ia Supernovae (SNeIa) and Gamma-Ray Burst (GRBs) occurrences.

By fitting the model parameters to these observational data sets, we extract the mean values for our proposed model. This thorough analysis enables us to align our model predictions with the observed data, resulting in a more accurate and well-defined cosmological framework. The integration of diverse observational data sets contributes to the reliability and precision of our constructed cosmological model. By studying different articles \citep{Colgain:2022tql,Colgain:2022rxy,Colgain:2023bge} they used both binning and not binning data by redshift highlighting the complexity of estimating cosmological parameters from observational data, especially at high redshifts. They show that while it is possible to recover Planck values in some cases, there may be non-Gaussian behavior in the probability density functions (PDFs) and potential challenges in accurately estimating parameters at higher redshifts. The choice of how to handle high-redshift data can impact the results and inferences drawn from cosmological models like $\Lambda$CDM.

\subsection{ Cosmic Chronometer (CC) Data Observations}
The CC method is a simple technique to measure the Hubble parameter, $H(z)$, as a function of redshift independent of the cosmological model. It relies on the relationship between time and redshift in an FLRW metric and is given by
\begin{equation*}
    H=-(1+z)^{-1} dz/dt.
\end{equation*} 
By obtaining $dt$ and $dz$ with sufficient precision, $H(z)$ can be measured in cosmology independently. While $\delta z /z \lesssim
0.001$ accuracy in redshift measurement is achievable, the main challenge lies in obtaining an estimate of the differential age evolution, $dt$. This requires the use of a `chronometer'. Passive stellar populations are ideal candidates as cosmic chronometers, as they evolve on much longer timescales compared to their age difference.
This relationship allows astronomers to infer the expansion rate of the Universe at different points in time, providing insight into the fundamental properties of the Universe. The methods discussed in the article  \citep{Koksbang:2021qqc} estimate the expansion rate of the Universe, and it highlights how their results can be affected by assumptions about the Universe's structure. 
R. Jimenez and A. Loeb \citep{Jimenez:2001gg} introduced a technique for retrieving HP data directly by calculating $dz/dt$ at a specific value of $z$. 

The primary advantage of the CC approach is its ability to directly estimate the expansion history of the Universe without the need for any prior cosmological assumptions. This characteristic makes it an ideal framework for rigorously testing various cosmological models.  Moreover, the CC method offers a cosmology-independent way to estimate the Universe's expansion history. However, its main challenge lies in systematic uncertainties, arising from  Stellar Population Synthesis (SPS) model choice, stellar metallicity estimation, Star Formation History (SFH) assumptions, and residual star formation. Addressing these uncertainties is crucial for accurate $H(z)$ measurements.

    \begin{table}
        \centering
        \caption{CC data yield values of $H(z)$ along with their associated 1$\sigma$ uncertainties, encompassing both systematic and statistical factors \protect\citep{Moresco:2020fbm} These values are presented in units of km/s/Mpc. The upper block represents the non-correlated data points of different surveys. The lower block presents correlated data points that have been taken into consideration, as elaborated in \protect\citep{Moresco:2020fbm}.}
         \begin{tabular*}{\columnwidth}{@{\extracolsep{\fill}}c c c}
        \hline\hline
         \multicolumn{3}{c}{Non-correlated data points}\\
         \hline
           Redshift $z$ & $Hz$ measure & Reference\\
            \hline
            $0.07$ & $69\pm19.6$ & \citep{Zhang:2012mp} \\ 
            $0.09$ & $69\pm12$ & \citep{Stern:2009ep} \\ 
            $0.12$ & $68.6\pm26.2$ & \citep{Zhang:2012mp} \\
            $0.17$ & $83\pm8$ & \citep{Stern:2009ep}\\
            $0.2$ & $72.9\pm29.6$ & \citep{Zhang:2012mp} \\ 
            $0.27$ & $77\pm14$ & \citep{Stern:2009ep}\\
            $0.28$ & $88.8\pm36.6$ & \citep{Zhang:2012mp} \\
            $0.4$ & $95\pm17$ & \citep{Stern:2009ep}\\
            $0.47$ & $89\pm34$ & \citep{Ratsimbazafy:2017vga}\\
            $0.48$ & $97\pm60$ & \citep{Stern:2009ep}\\
            $0.75$ & $98.8\pm33.6$ & \citep{Borghi:2021rft}\\
            $0.8$ & $113.1\pm28.5$ &\citep{Jiao:2022aep}\\  
             $0.88$ & $90\pm40$ & \citep{Stern:2009ep} \\
             $0.9$ & $117\pm23$ & \citep{Stern:2009ep}\\
             $1.3$ & $168\pm17$ & \citep{Moresco:2012jh}\\
             $1.43$ & $177\pm18$ & \citep{Stern:2009ep}\\
            $1.53$ & $140\pm14$ & \citep{Stern:2009ep} \\
            $1.75$ & $202\pm40$ & \citep{Stern:2009ep} \\
            $1.26$ & $135\pm65$ &\citep{Tomasetti:2023kek}\\
            \hline
            \multicolumn{3}{c}{Correlated data points}\\
         \hline
           Redshift $z$ & $Hz$ measure & Reference\\
        \hline
            $0.1791$& $75\pm4$ & \citep{Moresco:2012jh} \\ 
             $0.1993$ &  $75\pm5$  & \citep{Moresco:2012jh} \\
             $0.3519$ &$83\pm14$  &\citep{Moresco:2012jh}  \\
             $0.3802$  &$83\pm13.5$  &\citep{Moresco:2016mzx}  \\
             $0.4004$ & $77\pm10.2$ & \citep{Moresco:2016mzx} \\
             $0.4247$ & $87.1\pm11.2$ & \citep{Moresco:2016mzx} \\
             $0.4497$ & $87.1\pm11.2$ & \citep{Moresco:2016mzx}\\
             $0.4783$ & $80.9\pm9$ & \citep{Moresco:2016mzx}  \\
             $0.7812$ & $105\pm12$ & \citep{Moresco:2012jh} \\
             $0.5929$ & $104\pm13$ & \citep{Moresco:2012jh} \\
             $0.6797$ & $92\pm8$ & \citep{Moresco:2012jh} \\ 
             $0.8754$ & $125\pm17$ & \citep{Moresco:2012jh}  \\
             $1.037$ & $154\pm20$ & \citep{Moresco:2012jh} \\
             $1.363$ & $160\pm33.6$ & \citep{Moresco:2015cya} \\
             $1.965$ & $186.5\pm50.4$ & \citep{Moresco:2015cya}\\
            \hline\hline
        \end{tabular*}
        \label{tab:CC}
    \end{table}

Many recent studies have incorporated the CC data sets by analyzing their covariance matrix by \citep{Moresco:2020fbm}. The covariance matrix, denoted as $C_{ij}$, encompasses various contributions, including statistical errors ($C_{ij}^{stat}$), errors from young components ($C_{ij}^{young}$), dependence on the chosen model ($C_{ij}^{model}$), and uncertainties related to stellar metallicity ($C_{ij}^{stemet}$), the covariance matrix associated with the CC  method can be expressed as 
\begin{equation}
    C_{ij}=C_{ij}^{stat}+C_{ij}^{young}+C_{ij}^{model}+C_{ij}^{stemet}.\label{cov}
\end{equation}
Specifically, the model covariance, $C_{ij}^{model}$, can be further broken down into different components, each representing a distinct source of uncertainty. These components include uncertainties from the star formation history ($C_{ij}^{SFH}$), the initial mass function ($C_{ij}^{IMF}$), the stellar library ($C_{ij}^{Ste.Lib}$), and the stellar population synthesis model ($C_{ij}^{SPS}$). Thus, the model covariance is expressed as
\begin{equation}
    C_{ij}^{model}=C_{ij}^{SFH}+C_{ij}^{IMF}+C_{ij}^{Ste.Lib}+C_{ij}^{SPS}.
\end{equation}

 The recent work by \citep{Tomasetti:2023kek} aimed to derive a new constraint on the expansion history of the Universe by applying the cosmic chronometers method in the context of the VANDELS survey. Their focus is on studying the age evolution of high-redshift galaxies, using a full-spectral-fitting approach. The sample consisted of 39 massive and passive galaxies within the redshift range of 1 to 1.5.
By employing the cosmic chronometers method on the selected sample, they successfully obtained a new estimate of the Hubble parameter. The derived value for $H (z=1.26)$ was found to be $135\pm65$ km s$^{-1}$ Mpc$^{-1}$, taking into account both statistical and systematic errors. This finding provides crucial insights into the expansion rate of the Universe at that specific redshift.

In this work, we make use of 34 correlated and non-correlated points of the CC dataset, which has redshifts ranging from $0.07\leq z \leq 1.26$ and can be found in references \citep{Jimenez:2003iv,Chimento:2007da,Stern:2009ep,Moresco:2012jh,Zhang:2012mp,Moresco:2015cya,Moresco:2016mzx,Ratsimbazafy:2017vga,Borghi:2021rft,Jiao:2022aep}. To perform MCMC analysis, we need to evaluate the chi-square function of CC data, which is defined as follows
    \begin{gather}
    \chi^2_{CC/non\,cov}= \Delta A C_1^{-1}\Delta A^T\\
    \chi^2_{CC/cov}= \Delta A C_2^{-1}\Delta A^T.
    \end{gather}
    
Here, vector $A$ denotes a collection of CC data points, which have been computed as
\begin{equation}
    \Delta A_i=\left[H_{model}-H_{obs}(z_i)\right]_{1\times k}.
\end{equation}

  Here, $C_1^{-1}$ represents the inverse of the covariance matrix for the uncorrelated data points, which is explicitly defined as $C^{-1} = \left[1/\sigma^2 _H(z_i)\right]_{k\times k}$. In this equation, the index $i$ ranges from 1 to $k$, and $\sigma_H(z_i)$ corresponds to the respective errors associated with each data point in the observed CC data. Correspondingly, $C_2^{-1}$ stands for the inverse of the covariance matrix pertaining to correlated data points, and its precise formulation can be found in equation \eqref{cov}.  

 The total $\chi^2$ function for the CC observational data is as follows
 \begin{equation}
     \chi^2_{CC}=\chi^2_{CC/non \;cov}+\chi^2_{CC/ cov}.
 \end{equation}

\subsection{ Baryonic Acoustic Oscillations (BAO) Data Observations}
BAO is an important cosmological phenomenon that originated in the early Universe. During the early stages of cosmic evolution, acoustic density waves were imprinted in the primordial plasma due to the interaction between baryonic matter (ordinary matter) and radiation. These acoustic waves left a distinct signature in the density distribution of baryonic matter.
Over billions of years, as the Universe expanded and evolved, these acoustic waves froze into a characteristic length scale. This length scale, known as the BAO scale, serves as a standard ruler in the Universe's large-scale structure. It provides a unique and robust cosmic ruler that can be used to measure the expansion history of the Universe.

\begin{table}
        \centering
        \caption{Parameters and their measures for the corresponding redshift along with the associated 1$\sigma$ uncertainties obtained through BAO. The upper block represents the non-correlated data points of different surveys. The lower block presents correlated data points. The data points with a superscript `a' are considered from the galaxy clustering survey while data points with superscript `$b$' are considered from 6dFGS, SDSS, and WiggleZ surveys. For the set of measures with `$b$', the covariance has been taken into consideration, as elaborated in \protect\citep{Giostri:2012ek}.}
        \resizebox{\linewidth}{!}{
         \begin{tabular}{c c c c}
        \hline\hline
         \multicolumn{4}{c}{Non-correlated data points}\\
         \hline
         Redshift $z$ & Measurement & Parameter & Reference\\
         \hline
           $0.15$ & $664\pm25.0$ & $D_{v_{ratio}}$ & \citep{Ross:2014qpa} 
           \\ $0.24$ & $79.69\pm2.99$ & $H(z)$ & \citep{Gaztanaga:2008xz}\\
           $0.275$ & $0.1390\pm0.0037$ & $r_d/D_v$ & \citep{SDSS:2009ocz} \\ $0.3$ & $81.7\pm6.22$ & $H(z)$ & \citep{Oka:2013cba} \\
           $0.31$ & $78.18\pm4.74$ & $H(z)$ &\citep{BOSS:2016zkm}\\
           $0.32$ & $1264\pm25$ & $D_{v_{ratio}}$ & \citep{Tojeiro:2014eea} \\ $0.34$ & $83.8\pm3.66$ & $H(z)$ & \citep{Gaztanaga:2008xz} \\
           $0.36$ & $79.94\pm3.38$ & $H(z)$ & \citep{BOSS:2016zkm} \\
           $0.4$ & $82.04\pm2.03$ & $H(z)$ &\citep{BOSS:2016zkm}\\
           $0.43$ & $86.45\pm3.97$ & $H(z)$ & \citep{Gaztanaga:2008xz} \\
           $0.44$ & $84.81\pm1.83$ &$H(z)$ & \citep{Chuang:2013hya} \\
           $0.48$ & $87.79\pm2.03$ & $H(z)$ & \citep{Chuang:2013hya}\\
           $0.52$ & $94.35\pm2.64$& $H(z)$ & \citep{BOSS:2016zkm} \\
           $0.54$ & $9.212\pm0.41$ & $D_A/r_d$ & \citep{Seo:2012xy} \\
            $0.56$ & $93.34\pm2.3$ & $H(z)$ & \citep{BOSS:2016zkm}\\
            $0.57$ & $87.6\pm7.8$ & $H(z)$ & \citep{Seo:2012xy}  \\ 
            $0.57$ & $13.67\pm0.22$ & $D_v/r_d$ &\citep{Chuang:2013hya}\\
            $0.59$ & $98.48\pm3.18$ &$H(z)$ & \citep{BOSS:2016zkm} \\ 
            $0.64$ & $98.82\pm2.98$ & $H(z)$ & \citep{BOSS:2016zkm} \\ 
            $0.697$ & $1499\pm77$ & $D_{A_{ratio}}$ & \citep{Bautista:2017wwp}\\
            $0.72$ & $2353\pm63$ & $D_{v_{ratio}}$ & \citep{DES:2017rfo} \\ 
            $0.81$ & $10.75\pm0.43$ & $D_A/r_d$ &\citep{Hou:2020rse} \\ $0.874$ &$16780\pm109$ & $D_{A_{ratio}}$ & \citep{Ata:2017dya} \\
            $1.480$ & $13.23\pm0.47$ & $D_H/r_d$ &\citep{Busca:2012bu}\\
            $1.52$ & $3843\pm147.0$ & $D_{v_{ratio}}$ & \citep{BOSS:2016wmc}\\
            $2.3$ & $34188\pm1188$ & $H/r_d$ & \citep{BOSS:2016wmc} \\
            $2.33$ & $244\pm8$ & $H(z)$ & \citep{BOSS:2016wmc} \\ 
            $2.34$ & $222\pm8.5$ & $H(z)$ & \citep{BOSS:2014hwf} \\
            $2.34$ & $8.86\pm0.29$ & $D_H/r_d$ &\citep{BOSS:2016wmc}\\
            $2.36$ & $226\pm9.3$ & $H(z)$ & \citep{BOSS:2013igd} \\
            \hline
            \multicolumn{4}{c}{Correlated data points}\\
            \hline
             Redshift $z$ & measurement & parameter & Reference\\
              \hline
            $0.38^a$ & 1512.39 & $D_M(r_{d, \text{fid}}/r_d)$ &\citep{BOSS:2016wmc}\\
            $0.38^a$ & 81.2087 & $H(z)(r_d/r_{d, \text{fid}})$ &\citep{BOSS:2016wmc}\\
            $0.51^a$ & 1975.22 & $D_M(r_{d, \text{fid}}/r_d)$ &\citep{BOSS:2016wmc}\\
            $0.51^a$ & 90.9029 & $H(z)(r_d/r_{d, \text{fid}})$ &\citep{BOSS:2016wmc}\\
            $0.61^a$ & 2306.68 & $D_M(r_{d, \text{fid}}/r_d)$ &\citep{BOSS:2016wmc}\\
            $0.61^a$ & 98.9647 & $H(z)(r_d/r_{d, \text{fid}})$ &\citep{BOSS:2016wmc}\\
            $0.106^b$ & $0.336\pm0.015$ & $r_d(z_d)/D_v(z)$ &\citep{Beutler:2011hx}\\
            $0.2^b$ & $0.1905\pm0.0061$ & $r_d(z_d)/D_v(z)$ &\citep{SDSS:2009ocz}\\
            $0.35^b$ & $0.1097\pm0.0036$ & $r_d(z_d)/D_v(z)$ &\citep{SDSS:2009ocz}\\
            $0.44^b$ & $0.0916\pm0.0071$ & $r_d(z_d)/D_v(z)$ &\citep{Blake:2011en}\\
            $0.6^b$ & $0.0726\pm0.0034$ & $r_d(z_d)/D_v(z)$ &\citep{Blake:2011en}\\
            $0.73^b$ & $0.0592\pm0.0032$ & $r_d(z_d)/D_v(z)$ &\citep{Blake:2011en}\\
            \hline\hline
        \end{tabular}}
        \label{tab:BAO}
    \end{table}

Observations of the large-scale structure of the Universe allow us to detect the BAO peaks in the matter power spectrum. These peaks are related to characteristic scales imprinted in the early Universe. The standard cosmological model, cold dark matter (CDM), proposes that quantum fluctuations during inflation seed the initial matter distribution. After inflation, the Universe becomes radiation-dominated, with baryonic matter coupled to radiation through Thomson scattering. Sound waves emerge from overdensities, driven by radiation pressure. During recombination, photons decouple from baryons, and at the baryon drag epoch, the sound waves stall. As a result, each initial overdensity evolves into a centrally peaked perturbation surrounded by a spherical shell, and the radius of these shells is called the sound horizon, $r_d$.

The sound horizon $r_d$ can be used to determine the angular separation, $\delta_\theta$, and redshift separation, $\delta_z$, at a specific redshift $z$  respectively defined as follows
\begin{gather}
    \delta_\theta=r_d/(1+z)D_A(z),\\
    \delta_z=r_d/D_H(z).
\end{gather}
 Here, $D_A(z)$ is the angular diameter distance, and $D_H(z)$ is the Hubble distance at redshift $z$.
By selecting appropriate values of $r_d$ and by constraining cosmic parameters that determine $D_H(z)/r_d$ and $D_A(z)/r_d$, we can estimate $H(z)$. The ratios $D_H(z)/r_d$ and $D_A(z)/r_d$ depend on certain cosmic parameters, such as the matter density, dark energy density, and the equation of state of dark energy. These parameters determine the expansion rate of the Universe at different epochs. However, to maintain uniformity in the choice of $r_d$, we use $r_d = 147.74 Mpc$ obtained from the Planck collaboration to compute the observed Hubble parameter for the corresponding distance function of each survey by \citep{Stern:2009ep,Zhang:2012mp,Moresco:2015cya,Moresco:2016mzx,Ratsimbazafy:2017vga,Riess:2021jrx,Brout:2022vxf,Brout:2021mpj,Scolnic:2021amr}.

We employ the $\chi^2$ statistic to determine the mean parameter values and constraints for a given model. The majority of the data points utilized in our analysis are uncorrelated \citep{BOSS:2014hwf,Gaztanaga:2008xz,Blake:2012pj,Chuang:2013hya,Chuang:2012qt,Busca:2012bu,Oka:2013cba,BOSS:2013rlg,BOSS:2013igd,Bautista:2017zgn,BOSS:2016zkm,Ross:2014qpa,SDSS:2009ocz,Tojeiro:2014eea,Seo:2012xy,Anderson:2012sa,Bautista:2017wwp,DES:2017rfo,Hou:2020rse,Ata:2017dya,Busca:2012bu,BOSS:2016wmc}, so
\begin{equation}\label{eq:11}
\chi^2_{bao/ non \;cov}(\Phi)=\sum_{i=1}^{N}\left[\frac{(H_{model}(z_i,\Phi)-H_{obs}(z_i))^2}{\sigma_H^2(z_i)}\right],
\end{equation}
where $N$ denotes the number of non-correlated BAO data points, $H_{model}$ represents the theoretical value of the HP, $\Phi$ represents the model parameters, $H_{obs}$ represents the observed values of the HP from BAO analysis, and $\sigma_H(z_i)$ represents the respective error in the observed BAO data points.

Similarly, for the measurements of the dataset with subscript `$a$' and `$b$' in Table~\ref{tab:BAO} we have utilized correlated BAO radial measurements data obtained from galaxy survey and WiggleZ \citep{Blake:2011en}, 6dFGS \citep{Beutler:2011hx}, SDSS \citep{SDSS:2009ocz} surveys respectively. To compute the chi-square function for BAO $_{cov}$, we use the following equations
\begin{gather}
    \chi^2_{BAO/{cov^a}}= \Delta B^a C^{-1}_a\Delta {B^T}^a,\\
    \chi^2_{BAO/{cov^b}}= \Delta B^b C^{-1}_b\Delta {B^T}^b.
\end{gather}
Here vector $B$ is a collection of correlated BAO data points, and for the data points with a superscript `$a$' (of Table~\ref{tab:BAO}) this can be defined as
\[ \Delta B^a_i=
\begin{bmatrix}
    H_{model}-H_{obs}(z_i)\\
    D_{A_{model}}-D_{A_{obs}}(z_i)\\
\end{bmatrix},
\]
and, for data points  with a superscript `$b$' (of Table~\ref{tab:BAO}) vector $B$ is represented as 
\begin{equation}
    \Delta B^b_i=\left[\nu_{model}-\nu_{obs}(z_i)\right].
\end{equation}
Here $\nu=D_A(z^*)/D_v(z_i)$ for  volume-averaged angular diameter distance $D_v$. Moreover, $C_a^{-1}$ is the inverse covariance matrix $C_a$ for the correlated data points with superscript '$a$' (\tableautorefname~\ref{tab:BAO}), and we have \citep{Ryan:2019uor} 

\begin{widetext}

\[ C_a=
\begin{bmatrix}
624.707 & 23.729 & 325.332 & 8.34963 & 157.386 & 3.57778 \\
23.729 & 5.60873 & 11.6429 & 2.33996 & 6.39263 & 0.968056 \\
325.332 & 11.6429 & 905.777 & 29.3392 & 515.271 & 14.1013 \\
8.34963 & 2.33996 & 29.3392 & 5.42327 & 16.1422 & 2.85334 \\
157.386 & 6.39263 & 515.271 & 16.1422 & 1375.12 & 40.4327 \\
3.57778 & 0.968056 & 14.1013 & 2.85334 & 40.4327 & 6.25936 \\
\end{bmatrix}
\]

Similarly, the covariance matrix of the next set of correlated data points with superscript '$b$' (\tableautorefname~\ref{tab:BAO}) is given \citep{Giostri:2012ek} 
    
\[C_b=
\begin{bmatrix}
2.14085433 & 0.13114745 & 0.07552162 & 0.06306109 & 0.04998069 & 0.04075542 \\
0.13114718 & 0.39206383 & 0.1057302  & 0.0357534  & 0.02833729 & 0.02310691 \\
0.07552178 & 0.10573091 & 0.13466821 & 0.02058878 & 0.01631817 & 0.01330623 \\
0.06306109 & 0.03575347 & 0.02058873 & 0.44501308 & 0.08922379 & 0.01111097 \\
0.04998069 & 0.02833735 & 0.01631814 & 0.08922379 & 0.10890756 & 0.04924964 \\
0.04075542 & 0.02310696 & 0.0133062  & 0.01111097 & 0.04924964 & 0.0940857 \\
\end{bmatrix}
\]
\end{widetext}

Therefore, the total chi-squared function for the BAO data set is defined as follows
\begin{equation}
   \chi^2_{BAO}= \chi^2_{BAO/non\;cov}+\chi^2_{BAO/cov}.
\end{equation}

     \subsection{ Pantheon$^+$ ( P22 ) Data Observations}
The Pantheon$^+$ (P22) analysis represents an extension and improvement of the original Pantheon analysis. It incorporates a larger dataset of supernova type Ia (SNeIa), consisting of 1701 light curves from 1550 SNeIa, gathered from 18 different studies \citep{Riess:2021jrx,Malekjani:2023dky,Brout:2022vxf,Brout:2021mpj,Scolnic:2021amr}. These SNeIa have redshifts spanning the range of $0.001$ to $2.2613$. Notably, the P22 compilation includes 77 light curves corresponding to galaxies containing Cepheid distances.

Compared to the original Pantheon compilation by \citep{Pan-STARRS1:2017jku}, the P22 analysis introduces several significant enhancements. Firstly, it boasts a larger sample size, particularly at low redshifts (below 0.01). Additionally, the redshift range covered by P22 has been extended. Moreover, the analysis addresses various systematic uncertainties related to redshifts, peculiar velocities, photometric calibration, and intrinsic scatter models of Type Ia supernovae (SNeIa). In the past studies \citep{Pan-STARRS1:2017jku}, the EoS of dark energy and the Universe's expansion rate ($H_0$) have been analysed separately. Nevertheless, both parameters rely on almost the same SNeIa. The main reason for this separation is that determination of these two parameters is based on comparing SNeIa in different ranges of redshift.  For $H_0$, supernovae in nearby galaxies with redshifts below 0.01 are compared to those in the "Hubble flow" with redshifts between 0.023 and 0.15, excluding higher redshifts. In contrast, measurements of EoS usually involve supernovae up to redshifts around 2, but exclude those with redshifts below 0.01. Consequently, only supernovae within the range of 0.023 to 0.15 in redshift are commonly analyzed for both parameters.

Further, the latest improvements in the scale and calibration of Type Ia supernovae catalogues to constrain the specific nature and evolution of dark energy through its effect on the expansion history of the Universe, the Bayesian methodology is extended to comparing the scattering model of the data, testing for non-gaussianity in the Pantheon+ Hubble residuals is studied by \citep{Lovick:2023tnv,Dainotti:2024gca}.

The statistical and systematic covariance matrices are integrated and utilized to constrain cosmological models expressed as follows
\begin{equation}
    C_{stat}+C_{syst}=C_{stat+syst}
\end{equation}
 By minimizing the chi-square function, the model parameters can be constrained:
\begin{equation}\label{eq:13}
\chi^2_{SN}= \Delta D (C_{stat+syst})^{-1}\Delta D^T.
\end{equation}
 The vector $D$ represents the collection of 1701 supernova distance-modulus residuals, which have been computed as 
\begin{equation}
     \Delta D_i=\mu_i-\mu_{model}(z_i),\label{eq:14}
\end{equation}
where $\mu_i$ is the distance modulus of the $i^{th}$ SNeIa, $\mu_{model}(z_i)$ is the theoretical distance modulus at redshift $z_i$, and $\mu_i= m_i-M$, where $m_i$ is the apparent magnitude and $M$ is the fiducial magnitude of SNeIa. The theoretical distance modulus is given by
\begin{equation}\label{eq:15}
\mu_{model}(z,\Phi)=25+5\log_{10}\left(\frac{d_L(z,\Phi)}{1Mpc} \right).
\end{equation}

The luminosity distance, represented by the equation
\begin{equation}
    d_L(z,\Phi)=c(1+z)\int_0^z \frac{d\kappa}{H(\kappa)},\label{eq:16} 
\end{equation}
describes the distance between a supernova and an observer as a function of redshift $z$ and cosmological parameters $\Phi$, with the speed of light $c$ appearing as a constant. Although the parameters $M$ and $H_0$ are only degenerate in the analysis of Type Ia supernovae (SNeIa), limitations arise when considering the recently published SH0ES results, which relax both constraints. In our analysis, we take $M=-19.253$ which has been determined from SH0ES Cepheid host distances having great constraining power on $H_0$. In light of this, the distance  residual is expressed as
\begin{equation}\label{eq:17}
    \Delta D'_{i}=\begin{cases}
			\mu_i-\mu_i^{Cep}, & \text{if $i\in $ Cepheid hosts}\\
            \mu_i-\mu_{model}(z_i), & \text{otherwise}
		 \end{cases}
\end{equation}
 where $\mu_i^{Cep}$ refers to the Cepheid host-galaxy distance released by SH0ES. When calculating the covariance matrix for the Cepheid host-galaxy, it can be combined with the covariance matrix for SNeIa, as described by Equation (\ref{eq:13}). This combined covariance matrix, denoted by $C^{SN}_{stat+syst}+C^{cep}_{stat+syst}$, includes both statistical and systematic uncertainties from the P22 dataset and is used to constrain cosmological models in the analysis, as given by
\begin{equation}\label{eq:18}
    \chi^2_{P22}= \Delta D' (C^{SN}_{stat+syst}+C^{cep}_{stat+syst})^{-1}\Delta D'^T.
\end{equation}
To take into account the constraints from the combined CC, BAO, and P22 datasets, we utilize the total chi-square function, which is obtained by summing up the individual chi-square functions for each dataset
\begin{equation}
\chi^2_{T}=\chi^2_{CC}+\chi^2_{BAO}+\chi^2_{P22}.
\end{equation}

\begin{figure}
    \centering
    \includegraphics[width=0.75\linewidth]{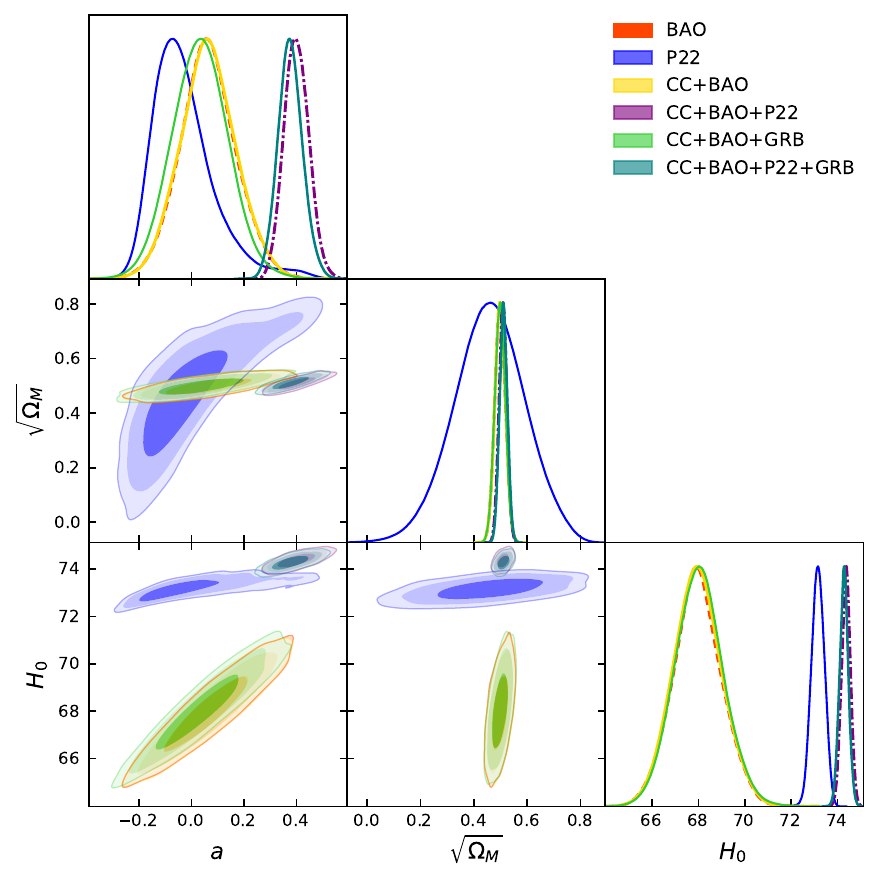}
    \caption{ A contour plot showcasing the model parameters $a$, $\sqrt{\Omega_M}$, and $H_0$, derived through $\chi^2$ analysis for the current model. The plot illustrates the results of a combined analysis involving diverse datasets, highlighting the confidence levels up to $3\sigma$.}
    \label{fig:param-I}
\end{figure}

\begin{figure}
    \centering
    \includegraphics[width=0.75\linewidth]{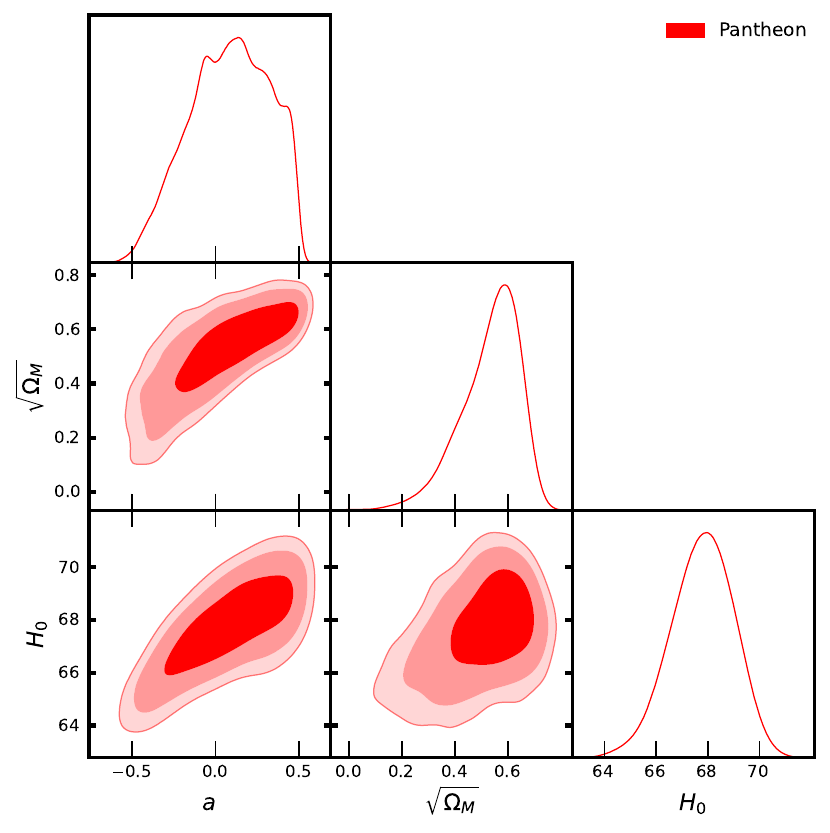}
    \caption{ A contour plot showcasing the model parameters $a$, $\sqrt{\Omega_M}$, and $H_0$, derived through $\chi^2$ analysis for the current model. The plot illustrates the result of 1048 samples of the Pantheon dataset, highlighting the confidence levels up to $3\sigma$.}
    \label{fig:panth}
\end{figure}

\subsection{Gamma-Ray Bursts Analysis}
Gamma-ray bursts (GRBs) represent the most powerful explosions in the Universe, exhibiting immense energy. These cosmic events remain observable even at incredibly high redshifts. As a result, GRBs hold the potential to study the expansion rate of the Universe and explore the characteristics of dark energy. To achieve this, it is crucial to accurately calibrate empirical correlations between the spectral and intensity properties of these bursts \citep{Dainotti:2017fhk,Dainotti:2017iqb,Parsotan:2022uur,Dainotti:2013cta,Dainotti:2013fra}.

The correlation between the rest-frame $\nu F_{\nu}$ spectrum peak energy, the observed photon energy of the peak spectral flux, $E_{p, i}$, and the isotropic-equivalent radiated energy, $E_{iso}$, initially discovered by \citep{Amati:2002ny}. and subsequently confirmed and extended through further observations, stands as one of the most fascinating and widely discussed pieces of observational evidence in the field of gamma-ray burst (GRBs) astrophysics. Here, is the considered correlation:

\begin{equation}
    \log\left(\frac{E_{iso}}{1 \,erg}\right)=b+a \log\left[\frac{E_{p,i}}{300\,kev}\right],\label{correlation}
\end{equation}

In this correlation, where $a$ and $b$ are constants, $E_{p, i}$ represents the spectral peak energy in the GRBs cosmological rest frame. It is related to the observer frame quantity, $E_p$, through the expression $E_{p, i}$ can be derived from the observer frame quantity $E_{p, i}=E_p (1+z)$, where $z$ is the redshift. This significant correlation not only imposes constraints on the model of the prompt emission during GRBs but also naturally suggests the potential use of GRBs as distance indicators. The isotropic equivalent energy, $E_{iso}$, can be determined using the bolometric fluence, Sbolo, as follows

\begin{equation}
    E_{iso}=4\pi d^2L(z,cp)S_{bolo}(1+z)^{-1}.
\end{equation}
The luminosity distance, $d_L$, is a crucial factor in this calibration process, representing the distance that light travels from the source (GRBs) to the observer. Additionally, $cp$ signifies the set of parameters that define the background cosmological model, such as the density of matter and dark energy in the Universe, the Hubble constant, and the curvature of space.

We analyzed a sample of 162 long GRBs ( Refer table 5 of the article \citep{Demianski:2016zxi}), where the redshift distribution spans a wide range, with values ranging from $0.03 \leq z \leq 9.3$. Notably, this redshift range extends well beyond the typical range observed for Type Ia supernovae (SNIa), which generally fall within the range of $z \leq 1.7$. This broad coverage of redshifts in our GRBs sample opens up new opportunities for studying cosmological phenomena and allows us to explore the Universe's properties at much higher redshifts.

Numerous correlations have been documented in the literature, many of which have found application in cosmological investigations. The researches conducted in \citep{Dainotti:2008vw,Dainotti:2010ki,Dainotti:2011yz,Dainotti:2015gva,Dainotti:2017fem} have extensively explored correlations pertaining to GRBs. \citep{Dainotti:2016iqn}  have extended this relation into three dimensions, forming what is known as the fundamental plane relation. This extension involves incorporating the prompt peak luminosity, $L_{Peak}$. Further extensions have been studied in \citep{Dainotti:2020jkj,Levine:2021gro,Dainotti:2022ked,Srinivasaragavan:2020isz}. 
\subsection{ Results}
 In this study, we employ a MCMC approach, to scan the redshift range of interest until the standard criteria for the convergence of the chains are reached.  In this work, we make use of different data sets, which include both correlated and non-correlated CC and BAO data sets, P22 data, and GRBs data. The contour plots in \figureautorefname~\ref{fig:param-I} depict the redshift ranges, that are consistent with the observational data at different confidence levels. These plots show regions where the model aligns well with the observed data, extending up to the $99.7\%$ confidence level.

The mean values and uncertainties of the model parameters from the analysis of the BAO data are $a=0.07\pm 0.11$, ${\Omega_M}=0.2693^{+0.0229}_ {-0.020}$, and $H_0=68.0\pm1.0$ km\;s$^{-1}$Mpc$^{-1}$ at 68\% Confidence Level (CL), and from the analysis of P22, we find $a=-0.027^{+0.069}_{-0.14}$, ${\Omega_M}=0.3481^{+0.011}_{-0.013}$, and  $H_0=73.16\pm 0.28$ km\;s$^{-1}$Mpc$^{-1}$ at 68\% CL. Furthermore, by combining CC and BAO data, the parametric values are $a=0.06\pm 0.11$, ${\Omega_M}=0.2693^{+0.023}_ {-0.020}$, and $H_0=68.0\pm1.0$ km\;s$^{-1}$Mpc$^{-1}$ at 68\% CL. Combining CC, BAO, and P22 data yields the parametric values $a=0.401\pm0.047$, ${\Omega_M}=0.2745^{+0.0234}_{-0.015}$, and $H_0=74.36\pm 0.19$ km\;s$^{-1}$Mpc$^{-1}$ (68\% CL), while the combined CC, BAO, and GRBs data result in $a=0.03\pm0.11$, ${\Omega_M}=0.2766^{+0.022}_ {-0.020}$, and $H_0=68.0\pm 1.1$ km\;s$^{-1}$Mpc$^{-1}$ (68\% CL). Finally, by combining all data (CC+BAO+P22+GRBs), we find the mean values with 1$\sigma$ error as $a=0.373\pm 0.046$, ${\Omega_M}=0.2766^{+0.2344}_ {-0.015}$, and $H_0=74.26\pm 0.19$ km\;s$^{-1}$Mpc$^{-1}$. 

These results demonstrate the compatibility of the model with the observational data from various sources. A notable increase in the parameter ${\Omega_M}$ is evident for the P22 dataset. Moreover, the Hubble parameter value is obtained around 74 for both combined CC+BAO+P22 and CC+BAO+P22+GRBs datasets. 

\begin{figure*}
     \centering
     \includegraphics[width=\linewidth]{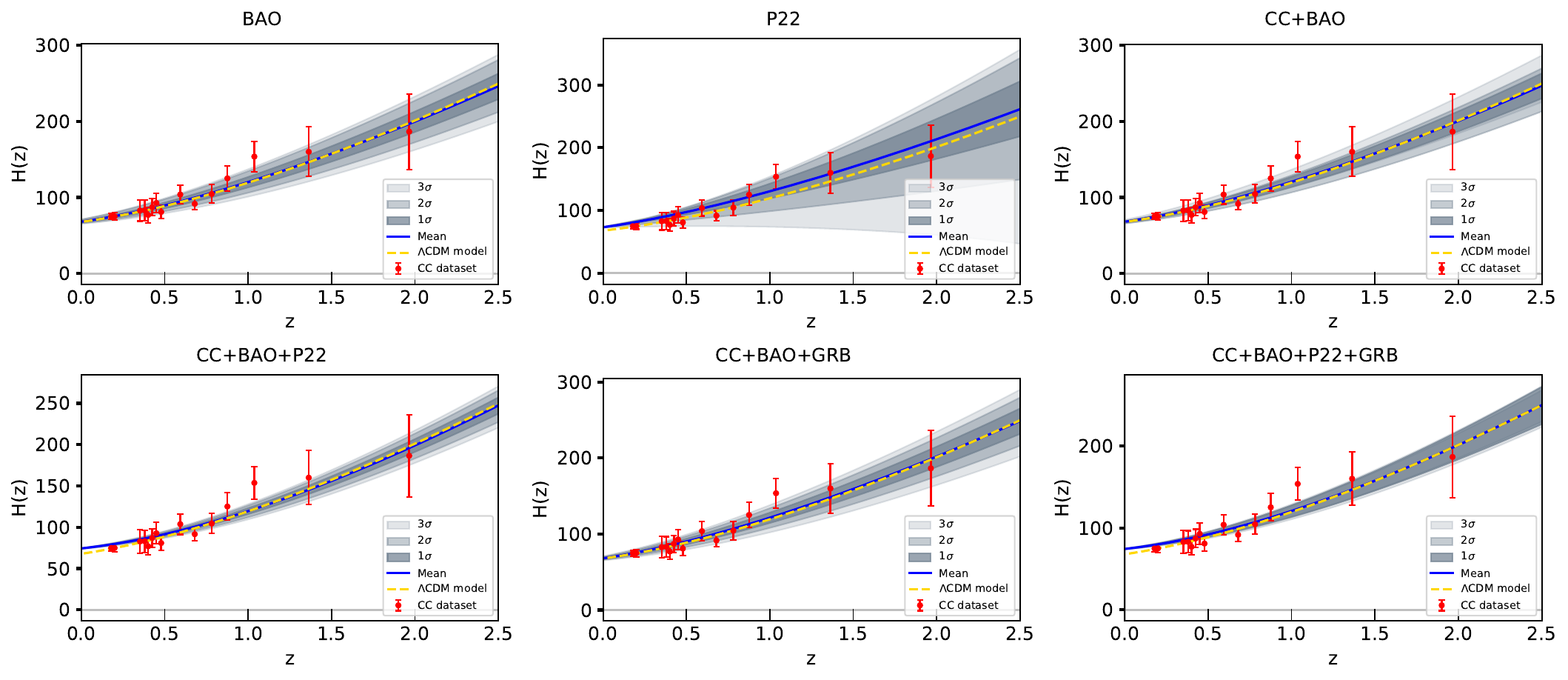}
     \caption{The plot illustrates the evolution of $H(z)$ up to a 3$\sigma$ confidence level (depicted by the gray shading) for the current model, incorporating diverse datasets. The plot features dots representing 34 CC data points, a yellow dashed curve for the standard $\Lambda$CDM model, and a solid curve within the shaded region depicting the evolution of $H(z)$ for the best-fit scenario.}
     \label{fig:errorplot}
 \end{figure*}

 \begin{figure*}
    \centering
    \label{fig:mu}{\includegraphics[width=0.99\linewidth]{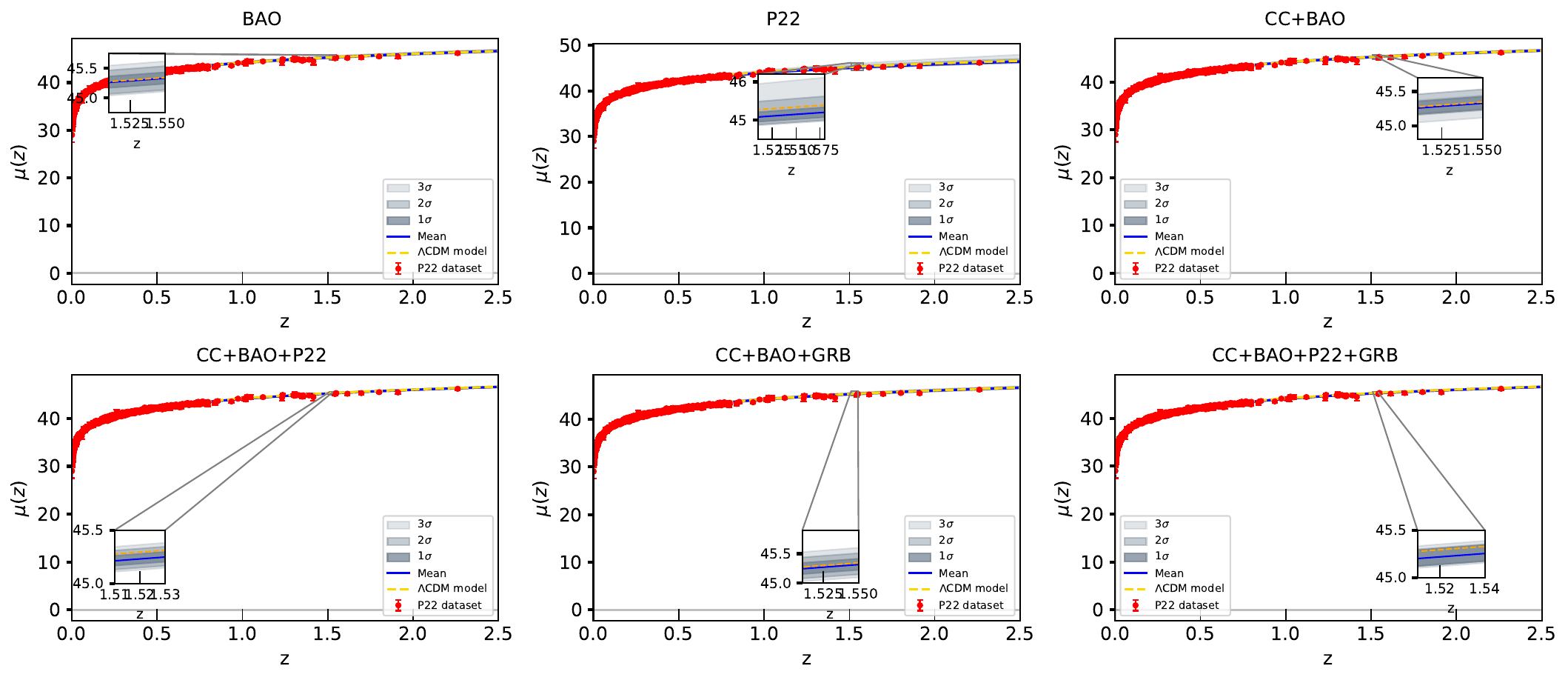}}
    \caption{Evolution of the distance modulus $\mu (z)$ is depicted, illustrating the mean values of model parameters resulting from the combined analysis of diverse datasets (depicted by the blue curve) with upto 3$\sigma$ confidence level (gray shaded region). The $\Lambda$CDM model ($\Omega_\lambda=0.7$, $\Omega_M=0.3$, and $H_0=67.8 kms^{-1}Mpc^{-1}$) is indicated by the yellow dashed line, serving as a reference for model comparison. The red dots on the plot represent the error bar plot comprising 1701 data points from the P22 compilation of Supernovae Type Ia dataset.}
    \label{fig:errmu}
\end{figure*}

\figureautorefname~\ref{fig:panth} showing contours with 3$\sigma$ confidence level derived from $\chi^2$ analysis for the current model, using 1048 samples of the Pantheon data sets only. Moreover, this plot serves as a cross-reference to a  \figureautorefname~\ref{fig:param-I}. Specifically to the P22 dataset which is depicted in dark blue color. These plots are plotted to present a comparative analysis of the former SNeIa sample with the updated one. We obtained $a=0.08^{+0.29}_ {-0.29}$, ${\Omega_M}=0.284^{+0.168}_ {-0.069}$, and $H_0=67.0^{+1.3}_ {-1.1}$ for 1048 SNeIa samples. Compared to 1048 SNeIa data points, the updated 1071 data points have higher constraining power of the Hubble parameter. The CMB data points present in 1048 SNeIa data are removed in the updated sample. The latter contains points attributed to Cephides that are responsible for obtaining a slightly higher value of constrained $H_0$.

In \figureautorefname~\ref{fig:errorplot}, we display the data for $H(z)$ along with their associated error bars for different data sets. The plotted blue line represents the mean theoretical curves derived from our cosmological model. To provide a comprehensive understanding of the uncertainties, we present grey-shaded regions that indicate the error bars at various confidence levels. Notably, we observe a remarkable agreement between the model predictions and the observed data, as the error bars closely align with the shaded regions. 

In \figureautorefname~\ref{fig:errmu}, we focus on the distance modulus function $\mu(z)$. Here, we present an error plot for the observed distance modulus of the 1701 SNeIa dataset. The blue line depicts the mean theoretical curves obtained from our cosmological model with constrained model parameters by different datasets. To account for uncertainties, the grey-shaded regions represent the error bars at an impressively high confidence level of up to 99.7$\%$. The agreement between the theoretical predictions and the observed distance modulus provides strong support for the accuracy and robustness of our model in describing the underlying cosmological processes.

\section{ Model Comparison: AIC and BIC Analysis}\label{model}
Within this section, our focus revolves around the comparison between a parametrized $H(z)$ model and a standard $\Lambda$CDM model (serving as the reference model). Additionally, we extend this comparison to encompass other parametrized models, including the EoS and Hubble parametrized models. To facilitate this comparison, we will employ widely recognized model selection statistics. The fundamental aim of a model selection statistic is to establish a delicate balance between a model's predictive power (often indicated by the number of free parameters it possesses) and its capacity to accurately conform to observed data. Therefore, in this work, we are utilizing two such statistics, the Akaike Information Criteria (AIC) \citep{Hakaike1974} and the Bayesian Information Criteria (BIC) \citep{Schwarz:1978tpv}, which have subsequently been quite widely applied to astrophysical problems. Applying them is relatively uncomplicated since they only demand the highest achievable likelihood within a specified model, as opposed to evaluating the likelihood across the entire redshift range.
The AIC is formulated as
\begin{equation}
    AIC = -2 \ln \mathcal{L}_{\text{max}} + 2k.
\end{equation}
In this context, $\mathcal{L}$ represents the likelihood (where $-2ln\mathcal{L}$ is frequently referred to as $\chi^2$, extending its applicability to non-Gaussian distributions), while $k$ denotes the count of parameters in the model. The subscript `max' indicates the requirement to determine parameter values that yield the utmost achievable likelihood within the model. The best model is the one that minimizes the AIC, denotes as $AIC^{*}$ and the models don't need to have a nested relationship. The disparity between AIC$^n$ and AIC$^{*}$, denoted by $\Delta$ AIC$^n$, is employed to gauge the degree of endorsement for the $n$th model. A $\Delta$ AIC$^n$ below 2 suggests the $n$th model is nearly on par with the best model. If $\Delta$ AIC$^n$ ranges between 4 and 7, the support for the $n$th model is notably weaker. A $\Delta$ AIC$^n$ surpassing 10 implies the $n$th model is unlikely to be the best.

The BIC is formulated as
\begin{equation}
   BIC = -2 \ln \mathcal{L}_{\text{max}} + k\;ln\;N,
\end{equation}

where $N$ represents the count of data points utilized in the fitting process. It is important to note that the BIC presupposes the independence and identical distribution of data points. However, the validity of this assumption depends on the specific dataset in question. For instance, it might not be suitable for cosmic microwave anisotropy data but could be appropriate for supernova luminosity-distance data. To identify the best model, we seek the model with the lowest BIC value, denoted as BIC$^{*}$. Analogous to $\Delta$AIC, we can calculate $\Delta$BIC$^n$ by subtracting the BIC value of the $n$th model from that of the best model (BIC$^{*}$). Among a set of models, the magnitude of $\Delta$BIC indicates evidence against the $n$th model as the best choice. A $\Delta$BIC$^n$ below 2 signifies weak evidence for the $n$th model compared to the best model. Values ranging from 2 to 6 suggest positive evidence against the $n$th model. When $\Delta$BIC$^n$ falls between 6 and 10, the evidence against the $n$th model is substantial. A $\Delta$BIC$^n$ exceeding 10 provides very strong evidence that the $n$th model is unlikely to be the best option.

Through observation, we note that the standard $\Lambda$CDM model exhibits the lowest AIC and BIC values (refer \tableautorefname~\ref{tab:tab_1}). Hence, we regard it as both the best and reference model. We intend to compare our model, along with the other two parametrized models to the best model.

Initially, we consider the HP parametrized model. The minimum $\chi^2$ value for our HP parametrized model stands at $1915.7182$, which is obtained through the combined CC, BAO, P22, and GRBs data set. By using statistical criteria, the AIC and BIC values corresponding to our model are determined as 1921.7182 and 1938.4279, respectively.

Further, In this study, We consider the CPL model, a well-known parametrization (EoS) proposed in \citep{Chevallier:2000qy,Linder:2002et}:

 \begin{equation}\label{eq:cpl}
\omega_{cpl}(z) = b + a\frac{z}{1+z},
\end{equation}

We proceed to carry out a statistical MCMC analysis using the collective data from CC, BAO, P22, and GRBs datasets. The 1$\sigma$ constrained values for $a$ and $b$ are determined to be $1.007^{+0.034}_{-0.034}$ and $-0.811^{+0.016}_{-0.016}$, respectively, accompanied by a minimum $\chi^2$ value of $1916.2266$.

Continuing our examination, we turn our focus to another parametrized model proposed by Abdulla Al Mamon \citep{AlMamon:2017tbm}, which we will refer to as the AAM model. The equation defines this model

\begin{equation}
H_{aam}(z) = H_0 \left[a + (1-a)(1+z)^b\right]^{3/2b}.
\end{equation}

Upon conducting an MCMC analysis on the AMM model, we derive the mean values within a 1$\sigma$ error range: $a = 0.783^{+0.015}_{-0.015}$ and $b = 3.31^{+0.18}_{-0.18}$, accompanied by a minimum $\chi^2$ value of 1919.4578.

After a careful examination of the outcomes, it becomes evident that our HP model demonstrates favorable results. This is supported by the fact that the $\Delta$ AIC value is less than two, indicating that our model is closely comparable to the best model ($\Lambda$CDM). Similarly, the other two models (CPL and AAM model) under investigation, display relatively elevated $\Delta$ AIC values compared to the $\Lambda$CDM model. Furthermore, the analysis of the BIC yields promising findings. The $\Delta$ BIC value of our HP model falls within the range of 2, highlighting its effectiveness similar to the best model ($\Lambda$CDM). Meanwhile,  the CPL and AAM models show $\Delta$ BIC results ranging between 2 to 6, implying evidence against these models when compared to the best model. These results indicate the consistency and statistical stability of our parameterized Model with respect to standard $\Lambda$CDM.

\section{Comparison of Methodologies}\label{dainotti}
\subsection{Dainotti relation: Gaussian likelihood analysis versus new likelihoods}
In their analysis `Pantheon' from \citep{Pan-STARRS1:2017jku} and `Pantheon+' from \citep{Scolnic:2021amr}, which accounted for both statistical and systematic uncertainties through covariance matrices. The former comprised 1048 sources within the redshift range of 0.01 to 2.26, drawn from various surveys including CfA1-4, Carnegie Supernova Project, Pan-STARRS1, Sloan Digital Sky Survey, Supernova Legacy Survey, and Hubble Space Telescope. The latter expanded the dataset to 1701 SNe Ia from 18 surveys across the range of 0.001 to 2.26 redshift, featuring an enhanced treatment of systematic uncertainties and a broader redshift span. Notably, the `Pantheon+' included 753 additional SNeIa compared to the `Pantheon', while the latter had 182 SNeIa not present in the former. These enhancements facilitated improved constraints on cosmological parameters. Both datasets were employed to assess the impact of these changes on the analysis of SNe Ia samples. However, we used expanded  1701 SNe Ia data from 18 surveys across the range of 0.001 to 2.2613 redshift only to asses the model to explore the accelerating Universe along with different datasets.

 \paragraph*{The Gaussianity assumption in the likelihood:} The study explores the Gaussianity assumption in the likelihood of SNe Ia distance moduli for both the Pantheon and Pantheon+ datasets within a flat $\Lambda$CDM model with $\Omega_M = 0.3$ and $H_0 = 70 kms^{-1} Mpc^{-1}$. They investigate normalized residuals ($\Delta_{\mu_{norm}}$) to account for statistical and systematic uncertainties in the covariance matrix. Although they tested different assumptions, such as varying $\Omega_M$, more specifically, they tested the two extreme cases $\Omega_M=0.1$ and $\Omega_M=1$. The Gaussianity tests consistently showed deviations from the Gaussian distribution for both datasets. These deviations persisted even under different cosmological assumptions, indicating that the results were independent of the specific cosmological model assumed.

The study employs Anderson-Darling \citep{darlo12} and Shapiro-Wilk \citep{shapiro23} normality tests, alongside skewness and kurtosis computations, to assess Gaussianity in the distributions of SNeIa. These tests detect deviations from Gaussian distribution, crucially considering even small deviations due to sample size. As sample sizes increase, these tests tend to reject normality, especially with large datasets like those in SNe Ia studies. Hence, skewness and kurtosis tests are included to provide additional insights into distribution characteristics. Skewness measures asymmetry, while kurtosis identifies extreme values in tails compared to a Gaussian distribution. By examining skewness and kurtosis, along with a combined `skewness+kurtosis' test, the study aims to evaluate the Gaussianity assumption in SNe Ia data.

\paragraph*{Fit with the new likelihoods:} 
The analysis extends beyond the initial tests' limitations, using Mathematica's Find Distribution tool to fit $\Delta_{\mu_{norm}}$ values. This tool compares distributions based on various statistical criteria like likelihood, BIC, AIC, and goodness-of-fit tests like Pearson $\chi^2$ and Cramer Von Mises tests. The top five fitting distributions for Pantheon and Pantheon+ samples are reported, along with their corresponding statistical test values in their work. Higher likelihood and lower BIC and AIC indicate better models (our model shows the same result as compared to this). Python computes BIC and AIC for easier interpretation. Distributions with larger p-values in goodness-of-fit tests are preferred. The first best-fit distribution is chosen for further analysis.

In their study, the researchers employed the logistic ($\mathcal{L}_{logistic}$) and Student’s ($\mathcal{L}_{student}$) likelihood functions to fit the flat $\Lambda$CDM model to both the Pantheon and Pantheon+ datasets. By doing so, they observed a notable reduction in the uncertainties associated with the parameters $\Omega_M $ and $H_0 = 70 $. Specifically, the uncertainties decreased by approximately 43\% and 41\% for $\Omega_M $, and by around 42\% and 33\% for $H_0 = 70 $ when using $\mathcal{L}_{logistic}$ and $\mathcal{L}_{student}$, respectively, compared to traditional Gaussian likelihoods. Their work highlights the efficacy of these alternative likelihood functions in providing more precise estimates of cosmological parameters, emphasizing the importance of selecting the appropriate likelihood function tailored to the characteristics of each SNeIa dataset (more info refer \citep{Dainotti:2024gca}).

\subsection{Calibration of the Dainotti relation through model-independent method}

We compare our work with \citep{Favale:2024lgp} as the authors have calibrated Dainotti's relation through a model-independent method using cosmic chronometer data. In our work, we analyze the parametrized Hubble model using 34 data points from a cosmic chronometer along with other data. In this section, we shall look into the corresponding work \citep{Favale:2024lgp}.

\paragraph*{Dainotti correlation:}
\begin{equation}
   \log L_X = C_0 + a \log T^*_X + b \log L_{\text{peak}}.
 \end{equation}
Here, $L_X$ represents the X-ray source rest-frame luminosity, $L_{peak}$ stands for the peak prompt luminosity, both in units of erg\, S$^{-1}$ denotes the characteristic time scale marking the end of the plateau emission, in seconds. 
The stability of the results obtained across various analyses conducted in the study reinforced the validity of using low-redshift data to calibrate the Dainotti relations. The analysis indicated a preference for a two-parameter relation, with the low-redshift data demonstrating the capability to identify a valuable set of standardizable candles. Specifically, the 20 GRBs selected within the redshift range of $0.553 \leq z \leq 1.96 $ were found to tightly adhere to the fundamental plane, leading to constraints on the 2D relation that align well with the physics governing this correlation. As a result, this subset of GRBs holds promise for future cosmological applications.

The study underscores the significance of finding novel distance indicators that are less susceptible to biases and systematics and can extend the range of applicability of the cosmic distance ladder to higher redshifts. This pursuit is crucial in cosmology and astrophysics, where the ability to obtain unbiased cosmological distances is essential, particularly amidst existing cosmological tensions.

Despite the promising results, the spread in observed luminosities of GRBs remains a significant challenge due to the varied nature of their origins, which could include a core collapse of massive stars or mergers of compact objects like neutron stars and black holes. Therefore, the reliability of these alternative probes and the underlying relations that highlight their intrinsic properties require thorough investigation, especially as data quality and quantity improve with upcoming surveys such as the SVOM and THESEUS missions.

However, so many correlations are discussed in the literature in our work we used the correlation mentioned in Equation (\ref{correlation}) with 162 long GRBs ranging from $0.03 \leq z \leq 9.3$.


\section{Model Kinematics of the Universe:  From Deceleration to Acceleration and the Jerk Evolution}\label{kinematics}

\par
\begin{figure*}
    \centering
    \label{fig:D-I}{\includegraphics[width=0.99\linewidth]{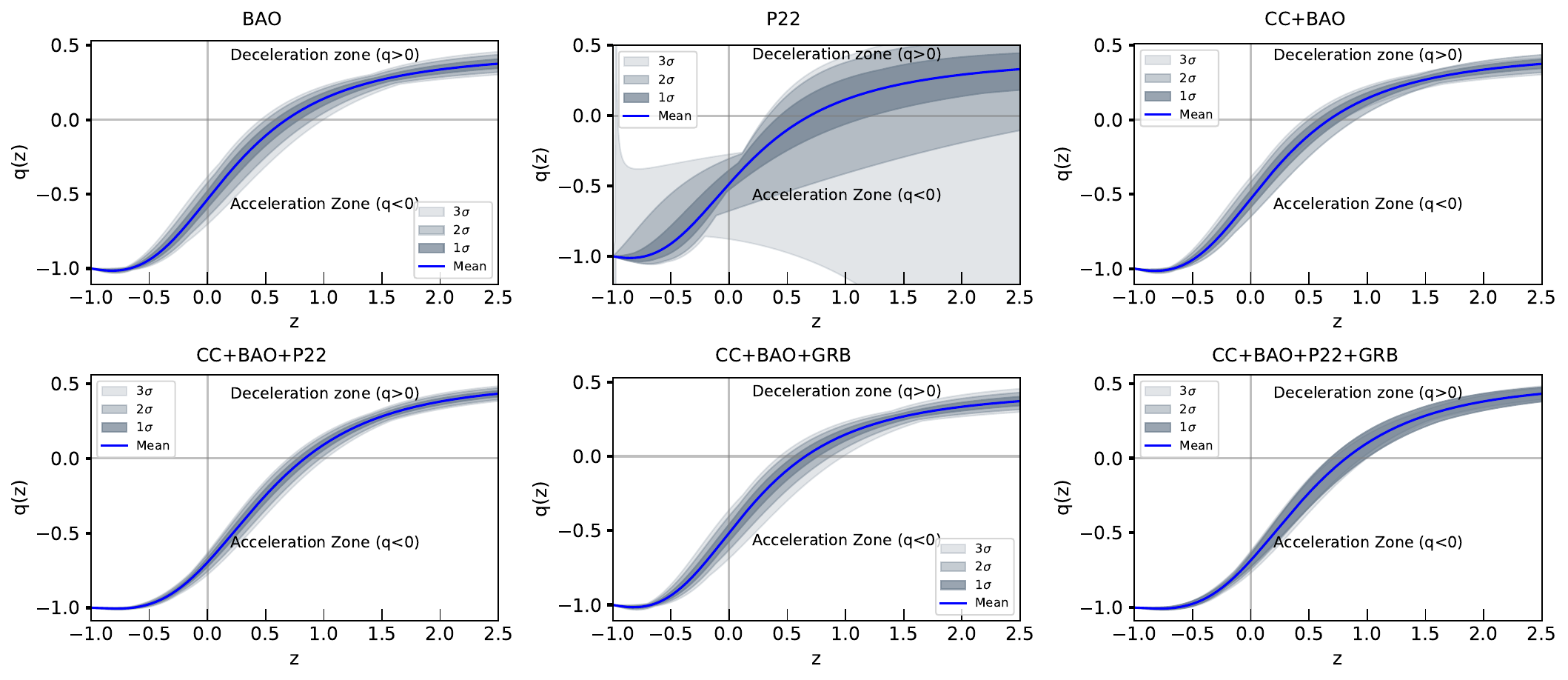}}
    \caption{Variation of $q$ with respect to redshift $z$, depicted using mean values of model parameters, resulting from a joint analysis of various datasets: CC, P22, CC+BAO, CC+BAO+P22, CC+BAO+GRBs, and CC+BAO+P22+GRBs (blue curve), with shaded gray regions indicating 68\%, 95\%, and 99.7\% confidence levels.}
    \label{fig:dcp}
\end{figure*}

 Observations reveal the Universe is undergoing accelerated expansion due to dark energy. However, for structure formation, a deceleration phase is needed at the onset of the matter-dominated era. The cosmological model must incorporate both deceleration and acceleration phases, with the $q$ playing a critical role. Positive $q$ signifies a decelerating phase driven by gravity, while negative $q$ indicates the current accelerating expansion caused by dark energy dominance. The transition from deceleration to acceleration occurred in the past, making $q$ essential for understanding the Universe's entire evolution. The transition from deceleration to acceleration in our model is illustrated by the evolution of the deceleration parameter in \figureautorefname~\ref{fig:dcp}. The combination of various cosmological data sets, including CC, BAO, P22, and GRBs, yields consistent results for the current $q$ ($q_0$). Thus, the obtained values with 1$\sigma$ error are approximately $q_0=-0.535^{+0.056}_{-0.059}$, $q_0=-0.485^{+0.097}_{-0.043}$, $q_0=-0.53^{+0.056}_{-0.059}$, $q_0=-0.697^{+0.028}_{-0.029}$, $q_0=-0.515^{+0.057}_{-0.058}$, and $q_0=-0.684^{+0.041}_{-0.043}$ across all the combinations of data: BAO, P22, CC+BAO, CC+BAO+P22, CC+BAO+GRBs, and CC+BAO+P22+GRBs respectively and are consistent with the values reported in previous studies \citep{Naik:2023yhl, Naik:2023gma,Basilakos:2011wm}.

The `transition redshift' is a critical concept in cosmology, marking the shift from a decelerating to an accelerating Universe. It is denoted as `$z_t$', representing the redshift value when dark energy's repulsive effects begin to counteract matter's deceleration. At this point, dark energy's density becomes comparable to matter, leading to the Universe's acceleration. Determining this value is vital for understanding the interplay between dark energy and matter throughout cosmic history.
Therefore, in our model, we have calculated the corresponding transition redshifts for various data sets, along with their 1$\sigma$ errors. The values are as follows: $z_t=0.677^{+0.110}_{-0.096}$, $z_t=0.695^{+0.528}_{-0.256}$, $z_t=0.671^{+0.109}_{-0.096}$, $z_t=0.836^{+0.065}_{-0.061}$, $z_t=0.656^{+0.109}_{-0.095}$, and $z_t=0.819^{+0.172}_{-0.135}$ for BAO, P22, CC+BAO, CC+BAO+P22, CC+BAO+GRBs, and CC+BAO+P22+GRBs data sets respectively. which is also in agreement with previous literature \citep{Naik:2023gma,Boughezal:2016wmq,Jesus:2017cyo}.

In the realm of cosmology, the EoS parameter plays a pivotal role in characterizing the behavior of dark energy, believed to drive the observed accelerated expansion of the Universe. Our examination of the EoS parameter within the framework of the proposed Hubble parameter model reveals quintessence behavior. This is supported by the obtained results: $w_t=-0.69^{+0.037}_{-0.039}$ (BAO), $w_t=-0.657^{+0.065}_{-0.028}$ (P22), $w_t=-0.687^{+0.037}_{-0.039}$ (CC+BAO), $w_t=-0.798^{+0.019}_{-0.02}$ (CC+BAO+P22), $w_t=-0.677^{+0.038}_{-0.039}$ (CC+BAO+GRBs), and $w_t=-0.79^{+0.027}_{-0.029}$ (CC+BAO+P22+GRBs), respectively. These values align with quintessence characteristics and are consistent with findings from prior studies, such as those reported in \citep{Sudharani:2023ywc, Linder:2002et}.

\begin{figure*}
    \centering
    \label{fig:Jerk1}{\includegraphics[width=0.99\linewidth]{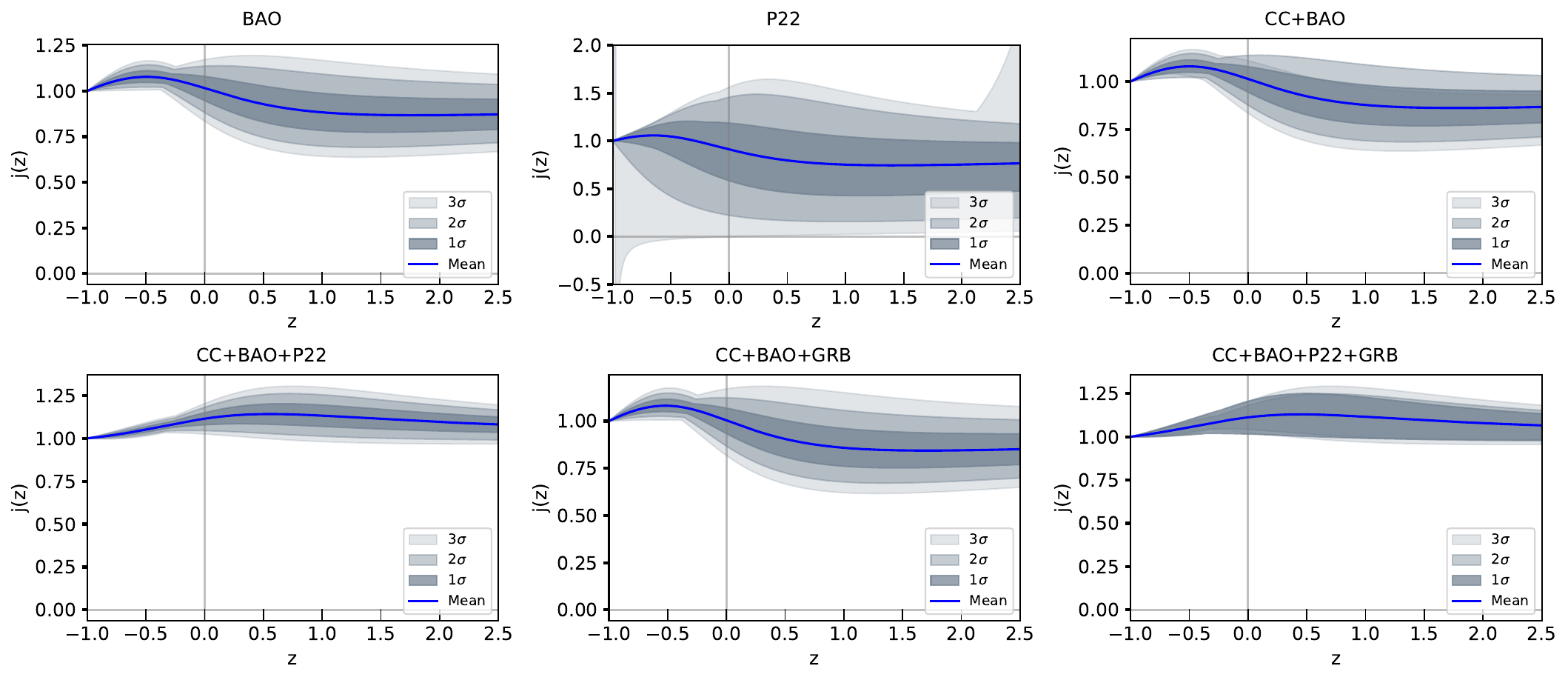}}
    \caption{The evolution of the jerk parameter's trajectory concerning redshift $z$ is depicted using the mean values( blue line) of model parameters with shaded gray region illustrates a confidence level of up to 3$\sigma$ while taking into consideration various sets of observational data.}
    \label{fig:Jerk}
\end{figure*}

Considering the transition from a decelerating phase to an accelerating phase in the history of the Universe, it becomes important to examine the third derivative of the scale factor $\mathcal{A}$. One convenient way to quantify this behavior is by using the dimensionless `jerk parameter (j)'. Additionally, we can express the jerk parameter as a function of redshift $z(t)$ using the deceleration parameter $q(z)$, and it is given by the equation:
\begin{equation}\label{eq:jerk}
j(z)=q(z)(2q(z)+1)+\frac{dq}{dz}(1+z).
\end{equation}
The equation incorporates both the value of $q(z)$ and its derivative with respect to redshift $z$ to determine $j$ at different epochs in the Universe's evolution. Also, its current value is denoted as `$j_0$'. The convenience of using the dimensionless jerk parameter lies in the fact that the $\Lambda$CDM model, simplifies to $j = 1$. The $\Lambda$CDM model serves as the baseline model, and by perturbing around it, the dimensionless jerk parameter provides a straightforward way to gauge deviations from this standard scenario. Thus, the current values of the jerk parameter for our constrained model with 1$\sigma$ CL are $j_0=1.016^{+0.064}_{-0.068}$, $j_0=0.912^{+0.275}_{-0.327}$, $j_0=1.013^{+0.064}_{-0.069}$, $j_0=1.115^{+0.035}_{-0.036}$, $j_0=1.004^{+0.065}_{-0.069}$, and $j_0=1.112^{+0.093}_{-0.097}$ across all the combinations of data: BAO, P22, CC+BAO, CC+BAO+P22, CC+BAO+GRBs, and CC+BAO+P22+GRBs respectively. These results are similar to the previous studies \citep{Boughezal:2016wmq,AlMamon:2018uby}. By observing \figureautorefname~\ref{fig:Jerk}, the jerk parameter for our parametrized model is slightly deviating from $j=1$. Therefore, our model is consistent with the standard model and is consistent with the value reported in previous studies. 

   \begin{table*}
    \caption{Comparative statistical analysis using the combined data sets for our model and other previously proposed parametrized models along with the best model ($\Lambda$CDM) .}
    \label{tab:tab_1}
     \setstretch{1.4}
	\centering
	\begin{tabularx}{\linewidth}{>{\centering\arraybackslash}X>{\centering\arraybackslash}X >{\centering\arraybackslash}X  >{\centering\arraybackslash}X >{\centering\arraybackslash}X  >{\centering\arraybackslash}X>{\centering\arraybackslash}X>{\centering\arraybackslash}X}
        \hline
        \hline
		{Models} &{$\chi^2_{min}$} & $a$ & $b$ & $AIC$ & $BIC$ &$\Delta AIC$ &$\Delta BIC$\\ 
		\hline
            $\Lambda$CDM- model & $1914.1108$ & $-0.584^{+0.022}_{-0.018}$ & $-0.723^{+0.015}_{-0.021}$ & $1920.1108$ & $1936.8205$ & $-$ & $-$\\
            
            HP-model & $1915.7182$ & $-0.719^{+0.03}_{-0.03}$ & $1.507^{+0.088}_{-0.088}$  & $1921.7182$ & $1938.4279$ & $1.921$& $1.6074$ \\
             
           CPL-model & $1916.2266$ & $-0.655^{+0.027}_{-0.027}$  & $0.737^{+0.043}_{-0.043}$ & $1922.2266$ & $1938.9363$ & $2.1158$& $2.1155$\\
            
           AAM-model & $1919.4578$ & $3.22^{+0.26}_{-0.33}$ & $2.89^{+0.20}_{-0.23}$  & $1925.4578$ & $1942.1675$ & $5.347$& $5.347$\\
            \hline
            \hline
	\end{tabularx}
    \end{table*}
\section{Diagnostic of the model}\label{diagnize}
The $Om(z)$ diagnostic proves to be a potent tool for discerning between diverse dark energy (DE) or cosmological models compared to the standard $\Lambda$CDM model. Introduced by \citep{Sahni:2008xx}, this diagnostic has since been extensively explored by numerous researchers. The function $Om(z)$ establishes a connection between the observed Hubble parameter, a measure of the Universe's expansion rate, and the density of matter within it. A constant $Om(z)$ value at any redshift signifies that the DE behaves akin to a cosmological constant. Conversely, varying $Om(z)$ with redshift suggests a dynamic nature of DE, indicating changes in its form over time. Additionally, the slope of $Om(z)$ serves to distinguish between two distinct types of dynamic DE models: quintessence and phantom. A positive slope in $Om(z)$ implies a phantom phase, while a negative slope suggests a quintessence phase.

\begin{figure}
    \includegraphics[width=0.9\linewidth]{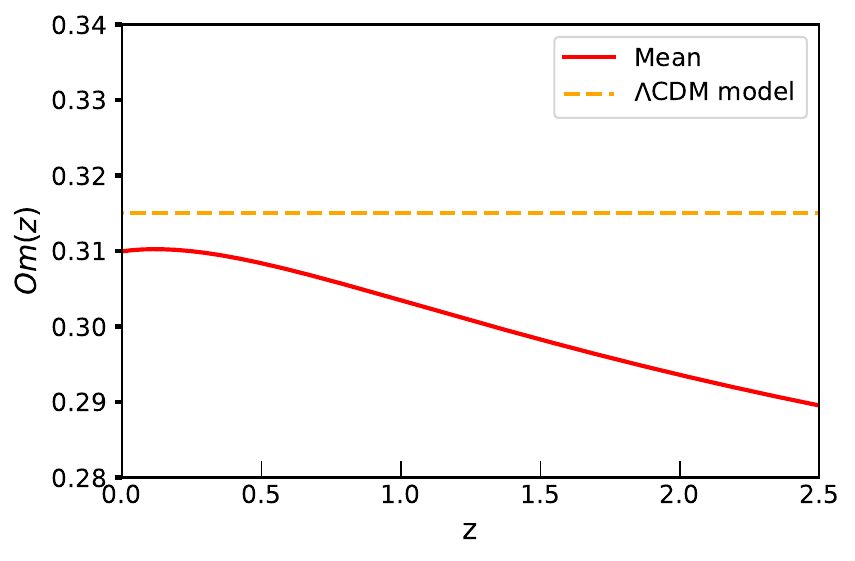}
    \caption{Evolution of the $Om(z)$ diagnostic across redshift \(z\) for the mean value of the constrained parameter, as inferred from combined datasets, indicates a quintessence phase.}
    \label{fig:om}
\end{figure}

In a Universe characterized by flat spatial geometry, the Om(z) diagnostic finds expression through the equation
\begin{equation}\label{Eq:Om}
    Om(z)=\frac{E^2(z)-1}{(1+z)^3-1},
\end{equation}
where $E(z)=H(z)/H_0$, employing the mean values of the constrained parameters obtained from the combined CC+BAO+P22+GRBs dataset, we depict the evolution of $Om(z)$ concerning $z$ in Figure~\ref{fig:om}. In the analysis, one can
observe that the mean value in Figure~\ref{fig:om} is less than that of the $\Lambda$CDM model. Consequently, the model falls into the quintessence region. We can observe that throughout the evolution the model shows quintessence behavior. For more
information, one may look at \citep{Escamilla-Rivera:2015odt}.

\begin{figure*}
     \centering
    \includegraphics[width=0.16\linewidth]{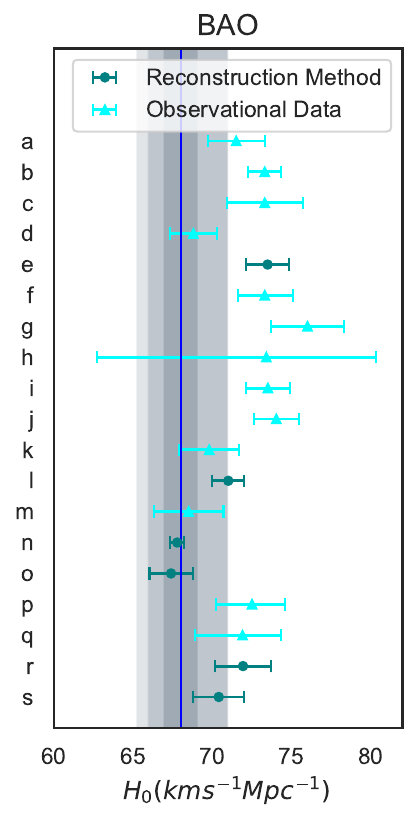}
    \includegraphics[width=0.16\linewidth]{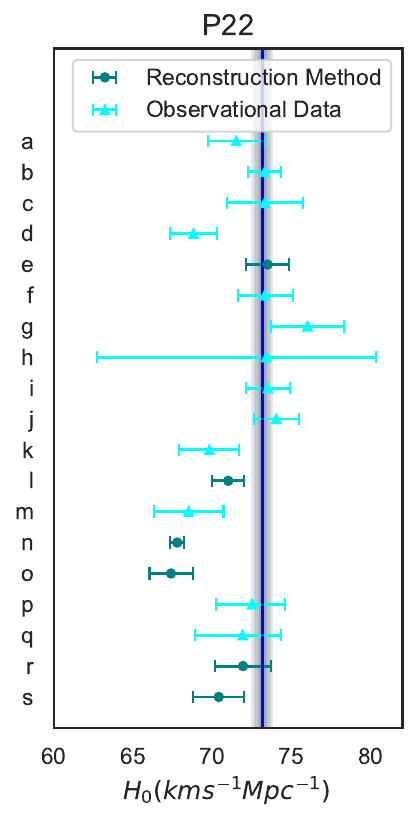}
    \includegraphics[width=0.16\linewidth]{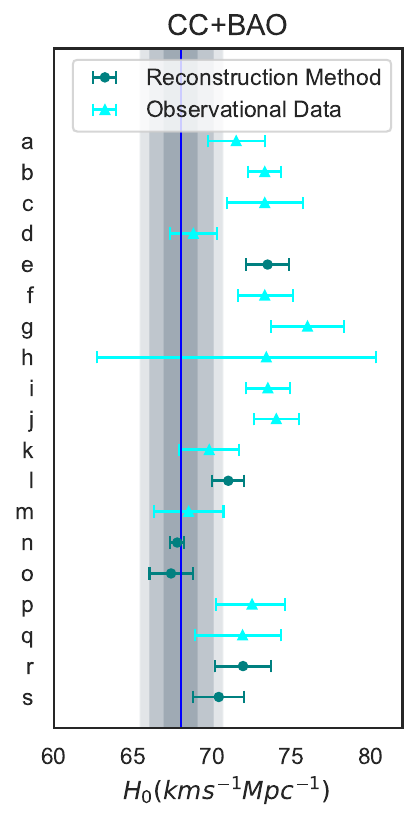}
    \includegraphics[width=0.16\linewidth]{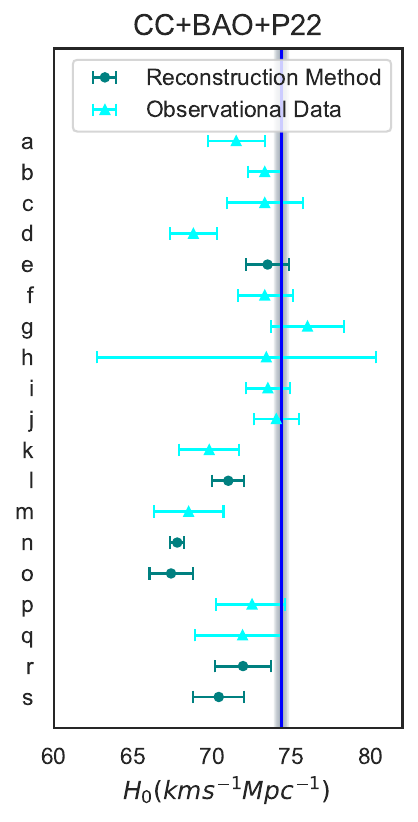}
    \includegraphics[width=0.16\linewidth]{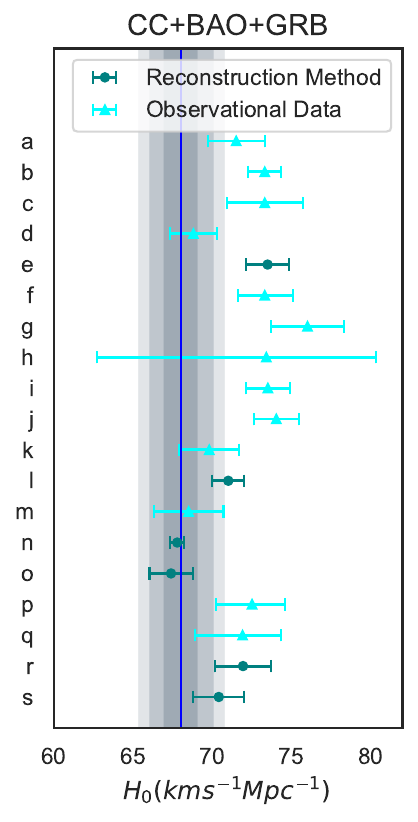}
    \includegraphics[width=0.16\linewidth]{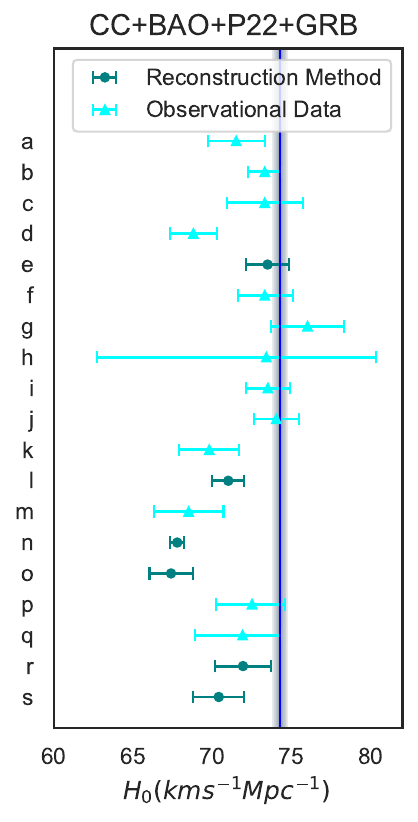} 
     \caption{
Comparing $H_0$ values from our current study to those from various sources. The solid blue line depicts the mean of the present $H_0$ value, while the shaded gray band shows error bars up to 3$\sigma$ confidence level. Alphabetic labels correspond to studies by the authors using both reconstructed methods and observational data: a) Anand et al. (2022), b) Riess et al. (2022), c) Blakeslee et al. (2021), d) Dutcher et al. (2021), e) Mehrabi et al. (2021), f) Wong et al. (2020), g) Kourkch et al. (2020), h) Gayathri et al. (2020), i) Reid et al. (2019), j) Riess et al. (2019), k) Freedman et al. (2019), l) Garza et al. (2019), m) D'Amico et al. (2019), n) Guo et al. (2019), o) Haridasu et al. (2018), p) Birrer et al. (2018), q) Bonvin et al. (2016), r) Tarrant et al. (2013), s) Campo et al. (2012). }\label{fig:comph}
   \end{figure*}

\section{Conclusions and Perspectives}\label{conclusion}
Researchers have extensively investigated the reconstruction approach for understanding cosmic evolution. This exploration involves two distinct methods: parametric and non-parametric reconstruction. Currently, there exists no universally accepted gravity theory capable of elucidating all aspects of the Universe. It is conceivable that both of these reconstruction approaches hold their individual advantages within this context. In this work, we introduce a cosmological model of the FLRW Universe utilizing the parametric approach. Notably, parametric methods have demonstrated their efficacy in elucidating the evolutionary trajectory of the Universe, encompassing its transition from early deceleration to subsequent acceleration. As a result, parameterization emerges as a promising avenue for effectively expounding upon and formulating future cosmological scenarios. Our central objective in this research is to reconstruct the  Hubble parameter, thereby delineating the progression of the contemporary universe.

In this work, we comprehensively explored how parameterized models illustrate cosmological dynamics. This investigation was facilitated by employing various observational data sets, including BAO, CC, P22 samples, and GRBs (Including both correlated and non-correlated data points). The utilization of Bayesian statistical inference techniques and MCMC methods to bound the model's parameters has enabled us to conduct precise data analysis and derive significant insights. The optimal fits obtained from this procedure were then employed to scrutinize the kinematic trends of the Universe. Our results reveal that the best-fit parameters of our models harmonize well with the observed data, suggesting that these proposed models present a credible portrayal of the Universe. The achievement in accurately bounding the model's parameters underscores the pivotal role of incorporating observational data and advanced statistical techniques to enhance our comprehension of the Universe's behavior. The work by \citep{Bargiacchi:2023jse, Dainotti:2023bwq} deals with cosmological probes together with GRBs, SNeIa, BAO, and Quasars. The recent studies by \citep{Favale:2024lgp} in which Cosmic Chronometers are applied together with GRBs.           
\par The values of the parameters acquired from statistical simulations exhibit notably symmetric uncertainties, where error bars symmetrically enclose the mean values. Nonetheless, for P22, the error bands at 2$\sigma$ and 3$\sigma$ levels display a slight asymmetry with a larger error range. Within these uncertainty bounds (2$\sigma$ and 3$\sigma$), the characteristics of the physical quantities exhibit more pronounced deviations. Consequently, to establish a reasonable range of uncertainty, we restrict our study to the 1$\sigma$ error.

Further in this study, we perform an analysis based on the Akaike Information Criterion (AIC) and Bayesian Information Criterion (BIC) by contrasting our proposed model with the $\Lambda$CDM model, considered the best model, as well as two other distinct models. The results obtained affirm the viability and effectiveness of our model when compared to the standard model, as outlined in the summarized \tableautorefname~\ref{tab:tab_1}. Additionally, we compare the Dainotti relation via Gaussian likelihood analysis versus new likelihoods and a model-independent approach to calibrating the Dainotti relation related to the datasets and correlation used in our analysis.

By exploring the dynamics of the model we put forth, which entails studying the transition from deceleration and conducting an analysis of the Jerk Parameter (j), we enhance our comprehension of how the Universe behaves and evolves. The current values for both the $q$ and the $j$, with uncertainties extended to a $3\sigma$ confidence level, are depicted separately in the \figureautorefname~\ref{fig:dcp} and \figureautorefname~\ref{fig:Jerk} respectively. Additionally, we examined the $Om(z)$ diagnostics to our model and is depicted in the Figure~\ref{fig:om}.

Moreover,
our investigation has placed constraints on both the model parameters and $H_0$. Notably, the values we derived for the Hubble constant are consistent with those obtained through other research endeavors that employed reconstruction approaches (both parametric and non-parametric) \citep{delCampo:2012ya,Roman-Garza:2018cxf,Haridasu:2018gqm,Mehrabi:2021cob,Tarrant:2013xka,Guo:2018ans}, as well as observational techniques such as the distance ladder method, TRGB technique, and H0LiCOW \citep{Riess:2019cxk,Freedman:2019jwv,Bonvin:2016crt,Birrer:2018vtm,Gayathri:2020mra, Blakeslee:2021rqi,Kourkchi:2020iyz,Wagle:2019vok,Wong:2019kwg,Riess:2021jrx,Anand:2021sum,DAmico:2019fhj,SPT-3G:2021eoc}. We illustrate our constrained $H_0$ findings in Figure \ref{fig:comph} and place them with the outcomes of preceding studies. Furthermore, the Hubble constant ($H_0$) tension is a recent and most interesting challenge in modern cosmology. Some studies are addressing these issues \citep{Giani:2023aor,Dainotti:2023yrk,Vagnozzi:2023nrq,Montani:2023ywn}.

The current study using CC+BAO+P22+GRBs yields a present-day Hubble parameter value of approximately $74.26km\;s^{-1}$Mpc$^{-1}$. Conversely, when considering CC+BAO+P22 alone, we obtain an $H_0$ value of $74.36km\;s^{-1}$Mpc$^{-1}$. Incorporating SHOES Cepheid Calibrated Pantheon data i.e., Pantheon+ data combined with other datasets (CC, BAO, GRB), has contributed to a slightly higher present-day Hubble value than initially anticipated. Notably, an observation emerges solely from the Pantheon+ data, we obtain the expected present Hubble value of around 73 units.

 Overall, our assessment of the parameterized Hubble parameter, guided by empirical data, indicates that the Universe is presently undergoing an accelerated phase. Importantly, our model aligns seamlessly with the observed outcomes, and these discoveries are poised to support the ongoing exploration of the Universe and its future trajectories. In forthcoming investigations, we intend to investigate different gravity theories via reconstructed kinematic models. We believe that, adopting a non-parametric approach to Universe reconstruction intriguing, as it has the potential to yield additional insights into cosmic evolution beyond the confines of parametric models. Exploring these research avenues has the potential to yield additional valuable results into the essence of the Universe and its evolution.
 
 \section*{Acknowledgements}
LS, NSK, and VV acknowledge DST, New Delhi, India, for supporting research facilities under DST-FIST-2019. The work of KB was partially supported by the JSPS KAKENHI Grant Number JP21K03547. LS and NSK acknowledge Kuvempu University for providing University General Fellowship (File no. KU: B.C.M-3/145/2023-24 Dated: 04/08/2023). We express our sincere gratitude to the esteemed referee for his/her valuable suggestions, which have greatly enhanced the quality and presentation of our research. The referees' insightful comments have played a crucial role in improving our work, and we are truly grateful for his/her contribution to this manuscript.

\section*{Data Availability}

There are no new data associated with this article.



\bibliographystyle{mnras.bst}
\bibliography{main} 




\bsp	
\label{lastpage}
\end{document}